\documentclass[sts,preprint]{imsart}
 
\usepackage[english]{babel}
\usepackage[utf8]{inputenc}
\usepackage{booktabs}
\usepackage{threeparttablex}
\usepackage{graphicx}
\usepackage[FIGTOPCAP]{subfigure}
\usepackage{epigraph} 
\usepackage[round]{natbib}
\usepackage{amsthm,amssymb,amsmath,natbib,bbm}
\usepackage[hyphens]{url}
\RequirePackage{hyperref}

\usepackage{comment}

\startlocaldefs

\numberwithin{equation}{section}

\newcommand{\real}{\mathbb{R}}

\newcommand{\bb}{\mathbf{b}}
\newcommand{\by}{\mathbf{y}}
\newcommand{\bA}{\mathbf{A}}
\newcommand{\bB}{\mathbf{B}}
\newcommand{\bQ}{\mathbf{Q}}
\newcommand{\bY}{\mathbf{Y}}

\newcommand{\bbeta}{{\boldsymbol \beta}}
\newcommand{\bLambda}{{\boldsymbol \Lambda}}
\newcommand{\bSigma}{{\boldsymbol \Sigma}}

\newcommand{\cA}{\mathcal{A}}
\newcommand{\cF}{\mathcal{F}}

\newcommand{\cN}{\mathcal{N}}

\newcommand{\mys}{{\rm s}}

\newcommand{\myE}{\mathbb{E}}
\newcommand{\myP}{\mathbb{P}}
\newcommand{\myQ}{\mathbb{Q}}
\newcommand{\myS}{{\rm S}}
\newcommand{\myT}{{\rm T}}

\newcommand{\dd}{\textup{d}}
\newcommand{\one}{\mathbbm{1}} 

\endlocaldefs

\begin{document}

\begin{frontmatter}

\title{Forecaster's Dilemma: Extreme Events and Forecast Evaluation}
\runtitle{Forecaster's Dilemma}

\begin{aug}

\author{\fnms{Sebastian} \snm{Lerch}\corref{}\ead[label=e1,text=sebastian.lerch@h-its.org]{sebastian.lerch@h-its.org}}\!\!,
\author{\fnms{Thordis L.} \snm{Thorarinsdottir}\ead[label=e2,text=thordis@nr.no]{thordis@nr.no}}\!\!,
\author{\fnms{Francesco} \snm{Ravazzolo}\ead[label=e3,text=francescoravazzolo@gmail.com]{francescoravazzolo@gmail.com}}\!\!
\and
\author{\fnms{Tilmann} \snm{Gneiting}\corref{}\ead[label=e4,text=tilmann.gneiting@h-its.org]{tilmann.gneiting@h-its.org}}

\address{Sebastian Lerch is Ph.D.~Student, Heidelberg Institute for
  Theoretical Studies (HITS), and Institute of Stochastics, Karlsruhe
  Institute of Technology, HITS gGmbH, Schloss-Wolfsbrunnenweg 35,
  69118 Heidelberg, Germany \printead{e1}.  Thordis L.~Thorarinsdottir
  is Senior Research Scientist, Norwegian Computing Center,
  P.O.~Box~114, Blin\-dern, 0314 Oslo, Norway \printead{e2}.
  Francesco Ravazzolo is Associate Professor, Free University of
  Bozen/Bolzano, Universitätsplatz 1, 39100 Bozen-Bolzano, Italy
  \printead{e3}.  Tilmann Gneiting is Group Leader, Heidelberg
  Institute for Theoretical Studies (HITS), and Professor of
  Computational Statistics, Institute of Stochastics, Karlsruhe
  Institute of Technology, HITS gGmbH, Schloss-Wolfsbrunnenweg 35,
  69118 Heidel\-berg, Germany \printead{e4}.}

\end{aug}

\runauthor{S.~Lerch, T.~L.~Thorarinsdottir, F.~Ravazzolo and T.~Gneiting}

\vspace{-7mm}

\begin{abstract} 
In public discussions of the quality of forecasts, attention typically
focuses on the predictive performance in cases of extreme events.
However, the restriction of conventional forecast evaluation methods
to subsets of extreme observations has unexpected and undesired
effects, and is bound to discredit skillful forecasts when the
signal-to-noise ratio in the data generating process is low.
Conditioning on outcomes is incompatible with the theoretical
assumptions of established forecast evaluation methods, thereby
confronting forecasters with what we refer to as the forecaster's
dilemma.  For probabilistic forecasts, proper weighted scoring rules
have been proposed as decision theoretically justifiable alternatives
for forecast evaluation with an emphasis on extreme events.  Using
theoretical arguments, simulation experiments, and a real data study
on probabilistic forecasts of U.S.~inflation and gross domestic
product (GDP) growth, we illustrate and discuss the forecaster's
dilemma along with potential remedies.
\end{abstract}


\end{frontmatter}

\vspace{-5mm}

\setlength{\epigraphwidth}{.53 \textwidth}
\epigraph{Quod male consultum cecidit feliciter, Ancus, \\
          Arguitur sapiens, quo modo stultus erat. \\
          Quod prudenter erat provisum, si male vortat, \\
          Ipse Cato (populo iudice) stultus erat.\footnotemark}{John Owen, 1607} 
          \footnotetext{\citet{Owen1607}, 216. \rule{0mm}{5mm}
          \textit{Sapientia duce, comite fortuna. In Ancum.} 
          English translation by Edith Sylla \citep{BernoulliSylla2006}: \\
          \textit{Because what was badly advised fell out happily, \\ 
                  Ancus is declared wise, who just now was foolish; \\
                  Because of what was prudently prepared for, if it turns out badly, \\
                  Cato himself, in popular opinion, will be foolish.}
         }

\section{Introduction}  \label{sec:introduction} 
 
Extreme events are inherent in natural or man-made systems and may
pose significant societal challenges.  The development of the
theoretical foundations for the study of extreme events started in the
middle of the last century and has received considerable interest in
various applied domains, including but not limited to meteorology,
climatology, hydrology, finance, and economics.  Topical reviews can
be found in the work of \citet{Gumbel1958}, \citet{EmbrechtsEtAl1997},
\citet{EasterlingEtAl2000}, \citet{Coles2001}, \citet{KatzEtAl2002},
\citet{BeirlantEtAl2004}, and \citet{AlbeverioEtAl2006}, among others.
Not surprisingly, accurate predictions of extreme events are of great
importance and demand.  In many situations distinct models and
forecasts are available, thereby calling for a comparative assessment
of their predictive performance with particular emphasis placed on
extreme events.

In the public, forecast evaluation often only takes place once an
extreme event has been observed, in particular, if forecasters have
failed to predict an event with high economic or societal impact.
Table~\ref{tab:media} gives examples from newspapers, magazines, and
broadcasting corporations that demonstrate the focus on extreme events
in finance, economics, meteorology, and seismology.  Striking examples
include the international financial crisis of 2007/08 and the L'Aquila
earthquake of 2009.  After the financial crisis, much attention was
paid to economists who had correctly predicted the crisis, and a
superior predictive ability was attributed to them.  In 2011, against
the protest of many scientists around the world, a group of Italian
seismologists was put on trial for not warning the public of the
devastating L'Aquila earthquake of 2009 that caused 309 deaths
\citep{Hall2011}.  Six scientists and a government official were found
guilty of involuntary manslaughter in October 2012 and sentenced to
six years of prison each. In November 2015, the scientists were
acquitted by the Supreme Court in Rome, whereas the sentence of the
deputy head of Italy's civil protection department, which had been
reduced to two years in 2014, was upheld.

At first sight, the practice of selecting extreme observations, while
discarding non-extreme ones, and to proceed using standard evaluation
tools appears to be a natural approach.  Intuitively, accurate
predictions on the subset of extreme observations may suggest superior
predictive ability.  However, the restriction of the evaluation to
subsets of the available observations has unwanted effects that may
discredit even the most skillful forecast available
\citep{DenrellFang2010, DiksEtAl2011, GneitingRanjan2011}.  In a
nutshell, if forecast evaluation proceeds conditionally on a
catastrophic event having been observed, always predicting calamity
becomes a worthwhile strategy.  Given that media attention tends to
focus on extreme events, skillful forecasts are bound to fail in the
public eye, and it becomes tempting to base decision-making on
misguided inferential procedures.  We refer to this critical issue as
the {\em forecaster's dilemma}.\footnote{Our notion of the {\em
forecaster's dilemma}\/ differs from a previous usage of the term in
the marketing literature by \citet{EhrmanShugan1995}, who investigated
the problem of influential forecasting in business environments.  The
forecaster's dilemma in influential forecasting refers to potential
complications when the forecast itself might affect the future
outcome, for example, by influencing which products are developed or
advertised.}

\begin{table}[t]

\footnotesize

\centering

\caption{Media coverage illustrating the focus on extreme events in
  public discussions of the quality of forecasts. A version of the
  table with links to the sources is provided in an online
  supplement. \label{tab:media}}

\begin{tabular}{lll} 
\toprule
Year & Headline & Source \\
\midrule
2008 & Dr.~Doom & The New York Times \\
2009 & How did economists get it so wrong? & The New York Times \\
2009 & He told us so & The Guardian \\
2010 & Experts who predicted US economy crisis see recovery & Bloomberg \\
     & in 2010 & \\
2010 & An exclusive interview with Med Yones - The expert who & CEO Q Magazine \\
     & predicted the financial crisis & \\
2011 & A seer on banks raises a furor on bonds & The New York Times \\
2013 & Meredith Whitney redraws `map of prosperity' & USA Today \\
\midrule
2007 & Lessons learned from Great Storm & BBC \\
2011 & Bad data failed to predict Nashville flood & NBC \\
2012 & Bureau of Meteorology chief says super storm `just blew up & The Courier-Mail \\
     & on the city'  & \\
2013 & Weather Service faulted for Sandy storm surge warnings & NBC \\
2013 & Weather Service updates criteria for hurricane warnings, & Washington Post \\
     & after Sandy criticism  & \\
2015 & National Weather Service head takes blame for forecast & NBC \\
     & failures & \\
\midrule
2011 & Italian scientists on trial over L'Aquila earthquake & CNN \\
2011 & Scientists worry over `bizarre' trial on earthquake & Scientific American \\
     & prediction  & \\
2012 & L'Aquila ruling: Should scientists stop giving advice? & BBC \\
\bottomrule
\end{tabular}

\end{table}

To demonstrate the phenomenon, we let $\cN(\mu,\sigma^2)$ denote the
normal distribution with mean $\mu$ and standard deviation $\sigma$
and consider the following simple experiment.  Let the observation $Y$
satisfy
\begin{equation}  \label{eq:sim1} 
Y | \, \mu \sim \cN(\mu,\sigma^2) 
\quad \textnormal{where} \quad \mu \sim \cN(0,1-\sigma^2).  
\end{equation}
Table \ref{tab:sim1a} introduces forecasts for $Y$, showing both the
predictive distribution, $F$, and the associated point forecast, $X$,
which we take to be the respective median or mean.\footnote{The
predictive distributions are symmetric, so their mean and median
coincide.  We use $X$ in upper case, as the point forecast may depend
on $\mu$ and $\tau$ and, therefore, is a random variable.}  The
perfect forecast has knowledge of $\mu$, while the unconditional
forecast is the unconditional standard normal distribution of $Y$.
The deliberately misguided extremist forecast shows a constant bias of
$\frac{5}{2}$.  As expected, the perfect forecast is preferred under
both the mean absolute error (MAE) and the mean squared error (MSE).
However, these results change completely if we restrict attention to
the largest 5\% of the observations, as shown in the last two columns
of the table, where the misguided extremist forecast receives the
lowest mean score.

In this simple example, we have considered point forecasts only, for
which there is no obvious way to abate the forecaster's dilemma by
adapting existing forecast evaluation methods appropriately, such that
particular emphasis can be put on extreme outcomes.  Probabilistic
forecasts in the form of predictive distributions provide a suitable
alternative.  Probabilistic forecasts have become popular over the
past few decades, and in various key applications there has been a
shift of paradigms from point forecasts to probabilistic forecasts, as
reviewed by \citet{TayWallis2000}, \citet{Timmermann2000},
\citet{Gneiting2008}, and \citet{GneitingKatzfuss2014}, among others.
As we will see, the forecaster's dilemma is not limited to point
forecasts and occurs in the case of probabilistic forecasts as well.
However, in the case of probabilistic forecasts extant methods of
forecast evaluation can be adapted to place emphasis on extremes in
decision theoretically coherent ways.  In particular, it has been
suggested that suitably weighted scoring rules allow for the
comparative evaluation of probabilistic forecasts with emphasis on
extreme events \citep{DiksEtAl2011, GneitingRanjan2011}.

\begin{table}[t]

\footnotesize

\centering

\caption{Forecasts in the simulation study, where the observation $Y$
  satisfies \eqref{eq:sim1} with $\sigma^2 = \frac{2}{3}$ being fixed.
  The mean absolute error (MAE) and mean squared error (MSE) for the
  point forecast $X$ are based on a sample of size 10\,000; the
  restricted versions rMAE and rMSE are based on the subset of
  observations exceeding 1.64 only.  The lowest value in each column
  is in bold.  \label{tab:sim1a}}
   
\begin{tabular}{lllcccc}
\toprule
Forecast      & Predictive Distribution & $X$ & MAE & MSE & rMAE & rMSE \\
\midrule
Perfect       & $\cN(\mu,\sigma^2)$ & $\mu$ & \bf{0.64} & \bf{0.67} & 1.35 & 2.12 \\
Unconditional & $\cN(0,1)$ & 0 \rule{0mm}{3.5mm} & 0.80 & 0.99 & 2.04 & 4.30 \\
Extremist \rule{0mm}{3.5mm} & $\cN(\mu + \frac{5}{2},\sigma^2)$ & $\mu + \frac{5}{2}$ 
              & 2.51 & 6.96 & \bf{1.16} & \bf{1.61} \\                
\bottomrule
\end{tabular}

\end{table}

The remainder of the article is organized as follows.  In Section
\ref{sec:theory} theoretical foundations on forecast evaluation and
proper scoring rules are reviewed, serving to analyse and explain the
forecaster's dilemma along with potential remedies.  In Section
\ref{sec:sim} this is followed up and illustrated in simulation
experiments.  Furthermore, we elucidate the role of the fundamental
lemma of Neyman and Pearson, which suggests the superiority of tests
of equal predictive performance that are based on the classical,
unweighted logarithmic score.  A case study on probabilistic forecasts
of gross domestic product (GDP) growth and inflation for the United
States is presented in Section \ref{sec:cs}.  The paper closes with a
discussion in Section \ref{sec:discussion}.

\section{Forecast evaluation and extreme events}  \label{sec:theory} 
 
We now review relevant theory that is then used to study and explain
the forecaster's dilemma.

\subsection{The joint distribution framework for forecast evaluation}  \label{sec:framework}
 
In a seminal paper on the evaluation of point forecasts,
\citet{MurphyWinkler1987} argued that the assessment ought to be based
on the joint distribution of the forecast, $X$, and the observation,
$Y$, building on both the {\em calibration-refinement factorization}, 
\[
[X,Y] \, = \, [X] \: [Y|X], 
\]
and the {\em likelihood-baserate factorization}, 
\[
[X,Y] \, = \, [Y] \: [X|Y]. 
\]
\citet{GneitingRanjan2013}, \citet{EhmEtAl2015}, and
\citet{StraehlZiegel2015} extend and adapt this framework to include
the case of potentially multiple probabilistic forecasts.  The joint
distribution of the probabilistic forecasts and the observation is
then defined on a probability space $(\Omega, \cA, \myQ)$, where the
elements of the sample space $\Omega$ can be identified with tuples
\[
(F_1, \ldots, F_k, Y), 
\]
the distribution of which is specified by the probability measure
$\myQ$.  The $\sigma$-algebra $\cA$ can be understood as encoding the
information available to forecasters.  The predictive distributions
$F_1, \ldots, F_k$ are cumulative distribution function (CDF)-valued
random quantities on the outcome space of the observation, $Y$.  They
are assumed to be measurable with respect to their corresponding
information sets, which can be formalized as sub-$\sigma$-algebras
$\cA_1, \ldots, \cA_k \subseteq \cA$.  The predictive distribution $F_i$
is {\em ideal}\/ relative to the information set $\cA_i$ if $F_i =
[Y|\cA_i]$ almost surely.  Thus, an ideal predictive distribution
makes the best possible use of the information at hand.  In the
setting of eq.~\eqref{eq:sim1} and Table \ref{tab:sim1a}, the perfect
forecast is ideal relative to knowledge of $\mu$, the unconditional
forecast is ideal relative to the empty information set, and the extremist 
forecast fails to be ideal.

Considering the case of a single probabilistic forecast, $F$, the
above factorizations have immediate analogues in this setting, namely, the 
calibration-refinement factorization 
\begin{equation}  \label{eq:CR}
[F,Y] \, = \, [F] \: [Y|F]
\end{equation}
and the likelihood-baserate factorization 
\begin{equation}  \label{eq:LB}
[F,Y] \, = \, [Y] \: [F|Y].    
\end{equation}
The components of the calibration-refinement factorization
\eqref{eq:CR} can be linked to the sharpness and the calibration of a
probabilistic forecast \citep{GneitingEtAl2007}.  Sharpness refers to
the concentration of the predictive distributions and is a property of
the marginal distribution of the forecasts only.  Calibration can be
interpreted in terms of the conditional distribution of the
observation, $Y$, given the probabilistic forecast, $F$.

Various notions of calibration have been proposed, with the concept of
auto-calibration being particularly strong.  Specifically, a probabilistic 
forecast $F$ is {\em auto-calibrated}\/ if
\begin{equation}  \label{eq:auto} 
[Y|F] = F 
\end{equation}
almost surely \citep{Tsyplakov2013}.  This property carries over to
point forecasts, in that, given any functional $\myT$, such as the
mean or expectation functional, or a quantile, auto-calibration
implies $\myT \left( [Y|F] \right) = \myT(F)$.  Furthermore, if the
point forecast $X = \myT(F)$ characterizes the probabilistic forecast,
as is the case in Table \ref{tab:sim1a}, where $\myT$ can be taken to
be the mean or median functional, then auto-calibration implies
\begin{equation}  \label{eq:auto.X} 
\myT \left( [Y|X] \right) = \myT \left( [Y|F] \right) = \myT(F) = X.    
\end{equation}
This property can be interpreted as unbiasedness of the point forecast
$X = \myT(F)$ that is induced by the predictive distribution $F$. 

Finally, a probabilistic forecast $F$ is {\em probabilistically
  calibrated}\/ if the probability integral transform $F(Y)$ is
uniformly distributed, with suitable technical adaptations in cases in
which $F$ may have a discrete component \citep{GneitingEtAl2007,
  GneitingRanjan2013}.  An ideal probabilistic forecast is necessarily
auto-calibrated, and an auto-calibrated predictive distribution is
necessarily probabilistically calibrated \citep{GneitingRanjan2013,
  StraehlZiegel2015}.

\begin{figure}[t]
    
\centering

\includegraphics[width = 0.60\textwidth]{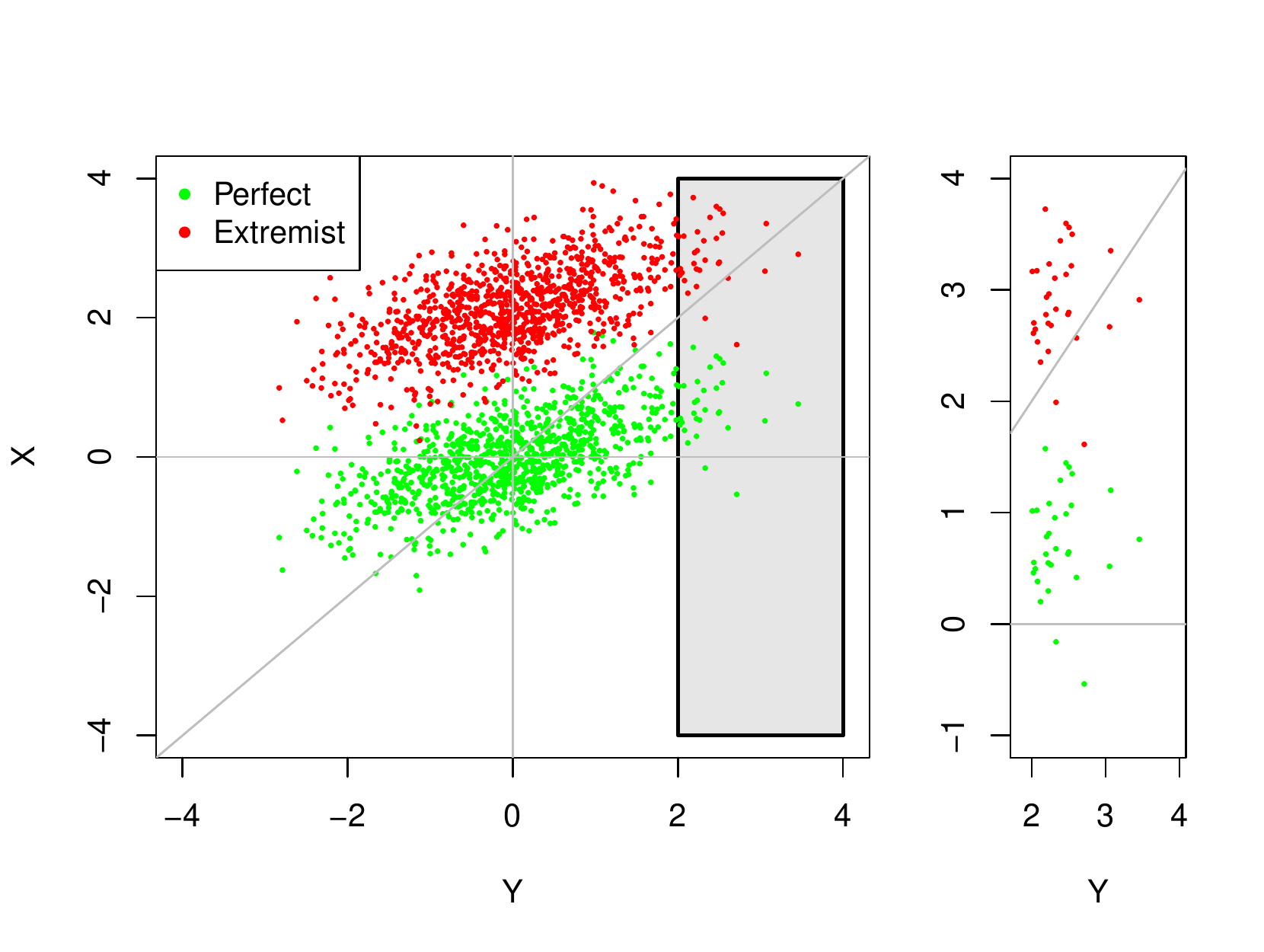}

\vspace{-3mm}

\caption{The sample illustrates the conditional distribution of the
  perfect forecast (green) and the extremist forecast (red) given the
  observation in the setting of eq.~\eqref{eq:sim1} and Table
  \ref{tab:sim1a}, where $\sigma^2 = \frac{2}{3}$.  The vertical
  stripe, which is enlarged at right, corresponds to cases where the
  respective point forecast exceeds a threshold value of
  2. \label{fig:LB}}

\end{figure}

In contrast, the interpretation of the second component $[F|Y]$ in the
likelihood-baserate factorization \eqref{eq:LB} is much less clear.
While the conditional distribution of the forecast given the
observation can be viewed as a measure of discrimination ability, it
was noted by \citet{MurphyWinkler1987} that forecasts can be perfectly
discriminatory although they are uncalibrated.  Therefore,
discrimination ability by itself is not informative, and forecast
assessment might be misguided if one stratifies by the realized value
of the observation.  To demonstrate this, we return to the simpler
setting of point forecasts and revisit the simulation example of
eq.~\eqref{eq:sim1} and Table \ref{tab:sim1a}, with $\sigma^2 =
\frac{2}{3}$ being fixed.  Figure \ref{fig:LB} shows the perfect
forecast, the deliberately misspecified extremist forecast, and the
observation in this setting.  The bias of the extremist forecast is
readily seen when all forecast cases are taken into account.  However,
if we restrict attention to cases where the observation exceeds a high
threshold of 2, it is not obvious whether the perfect or the extremist
forecast is preferable.\footnote{To provide analytical results,
$X_\textnormal{perfect} | Y = y \sim \cN \! \left( (1-\sigma^2)y,
\sigma^2(1-\sigma^2) \right)$ and $X_\textnormal{extremist} | Y = y
\sim \cN \! \left( (1-\sigma^2)y + \frac{5}{2}, \sigma^2(1-\sigma^2)
\right)$.}
     
In this simple example, we have seen that if we stratify by the value
of the realized observation, a deliberately misspecified forecast may
appear appealing, while an ideal forecast may appear flawed, even
though the forecasts are based on the same information set.
Fortunately, unwanted effects of this type are avoided if we stratify
by the value of the forecast.  To see this, note that ideal predictive
distributions and their induced point forecasts satisfy the
auto-calibration property \eqref{eq:auto} and, subject to conditions,
the unbiasedness property \eqref{eq:auto.X}, respectively.

\subsection{Proper scoring rules and consistent scoring functions}  \label{sec:scores}
 
In the previous section we have introduced calibration and sharpness
as key aspects of the quality of probabilistic forecasts.  Proper
scoring rules assess calibration and sharpness simultaneously and play
key roles in the comparative evaluation and ranking of competing
forecasts \citep{GneitingRaftery2007}.  Specifically, let $\cF$ denote
a class of probability distributions on $\Omega_Y$, the set of
possible values of the observation $Y$.  A {\em scoring rule}\/ is a
mapping $\myS : \cF \times \Omega_Y \longrightarrow \real \cup \{
\infty \}$ that assigns a numerical penalty based on the predictive
distribution $F \in \cF$ and observation $y \in \Omega_Y$.  Generally,
we identify a predictive distribution with its CDF.  A scoring rule is
{\em proper}\/ relative to the class $\cF$ if
\begin{equation}  \label{eq:proper}
\myE_G \, \myS(G,Y) \leq \myE_G \, \myS(F,Y)
\end{equation} 
for all probability distributions $F, G \in \cF$.  It is {\em strictly
proper}\/ relative to the class $\cF$ if the above holds with equality
only if $F = G$.  In what follows we assume that $\Omega_Y = \real$.
Scoring rules provide summary measures of predictive performance, and
in practical applications, competing forecasting methods are compared
and ranked in terms of the mean score over the cases in a test set.
Propriety is a critically important element that encourages honest and
careful forecasting, as the expected score is minimized if the quoted
predictive distribution agrees with the actually assumed, under which
the expectation in \eqref{eq:proper} is computed.

The most popular proper scoring rules for real-valued quantities are
the {\em logarithmic score}\/ (LogS), defined as
\begin{equation}  \label{eq:LogS}
\textnormal{LogS}(F,y) = - \log f(y),
\end{equation}
where $f$ denotes the density of $F$ \citep{Good1952}, which applies
to absolutely continuous distributions only, and the {\em continuous
ranked probability score}\/ (CRPS), which is defined as
\begin{equation}  \label{eq:CRPS}
\textnormal{CRPS}(F,y) =  
\int_{-\infty}^{\infty} \left( F(z) - \one \{ y \leq z \} \right)^2 \: \dd z
\end{equation}
directly in terms of the predictive CDF \citep{MathesonWinkler1976}.
The CRPS can be interpreted as the integral of the proper
Brier score \citep{Brier1950, GneitingRaftery2007}, 
\begin{equation}  \label{eq:BS}
\textnormal{BS}_z(F,y) = \left( F(z) - \one \{ y \leq z \} \right)^2,
\end{equation}
for the induced probability forecast for the binary event of the
observation not exceeding the threshold value $z$.  Alternative
respresentations of the CRPS are discussed in
\citet{GneitingRaftery2007} and \citet{GneitingRanjan2011}.

The quality of point forecasts is typically assessed by means of a
{\em scoring function}\/ $\mys(x,y)$ that assigns a numerical score
based on the point forecast, $x$, and the respective observation, $y$.
As in the case of proper scoring rules, competing forecasting methods
are compared and ranked in terms of the mean score over the cases in a
test set.  Popular scoring functions include the squared error,
$\mys(x,y) = (x - y)^2$, and the absolute error, $\mys(x,y) = |x -
y|$, for which we have reported mean scores in Table \ref{tab:sim1a}.

To avoid misguided inferences, the scoring function and the
forecasting task have to be matched carefully, either by specifying
the scoring function ex ante, or by employing scoring functions that
are {\em consistent}\/ for a target functional $\myT$, relative to the
class $\cF$ of predictive distributions at hand, in the technical
sense that
\[ 
\myE_F \, \mys(\myT(F),Y) \leq \myE_F \, \mys(x,Y) 
\] 
for all $x \in \real$ and $F \in \cF$ \citep{Gneiting2011}.  For
instance, the squared error scoring function is consistent for the
mean or expectation functional relative to the class of the
probability measures with finite first moment, and the absolute error
scoring function is consistent for the median functional.
 
Consistent scoring functions become proper scoring rules if the point
forecast is chosen to be the Bayes rule or optimal point forecast
under the respective predictive distribution.  In other words, if the
scoring function $\mys$ is consistent for the functional $\myT$, then
\[   
\myS(F,y) = \mys(\myT(F),y)
\]
defines a proper scoring rule relative to the class $\cF$.  For
instance, squared error can be interpreted as a proper scoring rule
provided the point forecast is the mean of the respective predictive
distribution, and absolute error yields a proper scoring rule if
the point forecast is the median of the predictive distribution.

\subsection{Understanding the forecaster's dilemma}  \label{sec:understanding}

We are now in the position to analyze and understand the forecaster's
dilemma both within the joint distribution framework and from the
perspective of proper scoring rules.  While there is no unique
definition of extreme events in the literature, we follow common
practice and take extreme events to be observations that fall into the
tails of the underlying population.  In public discussions of the
quality of forecasts, attention often falls exclusively on cases with
extreme observations.  As we have seen, under this practice even the
most skillful forecasts available are bound to fail in the public eye,
particularly when the signal-to-noise ratio in the data generating
process is low.  In a nutshell, if forecast evaluation is restricted
to cases where the observation falls into a particular region of the
outcome space, forecasters are encouraged to unduly emphasize this
region.

Within the joint distribution framework of Section
\ref{sec:framework}, any stratification by, and conditioning on, the
realized values of the outcome is problematic and ought to be avoided,
as general theoretical guidance for the interpretation and assessment
of the resulting conditional distribution $[F|Y]$ does not appear to
be available.  In view of the likelihood-baserate factorization
\eqref{eq:LB} of the joint distribution of the forecast and the
observation, the forecaster's dilemma arises as a consequence.
Fortunately, stratification by, and conditioning on, the values of a
point forecast or probabilistic forecast is unproblematic from a
decision theoretic perspective, as the auto-calibration property
\eqref{eq:auto} lends itself to practical tools and tests for
calibration checks, as discussed by \citet{GneitingEtAl2007},
\citet{HeldEtAl2010}, and \citet{StraehlZiegel2015}, among others.
 
From the perspective of proper scoring rules,
\citet{GneitingRanjan2011} showed that a proper scoring rule $S_0$ is
rendered improper if the product with a non-constant weight function
$w(y)$ is formed.  Specifically, consider the weighted scoring rule
\begin{equation}  \label{eq:improper} 
\myS(F,y) = w(y) \, \myS_0(F,y).  
\end{equation} 
Then if $Y$ has density $g$, the expected score $\myE_g \myS(F,Y)$  
is minimized by the predictive distribution $F$ with density
\begin{equation}  \label{eq:hedge} 
f(y) = \frac{w(y)g(y)}{\int w(z)g(z) \, \dd z},  
\end{equation} 
which is proportional to the product of the weight function, $w$, and
the true density, $g$.  In other words, forecasters are encouraged to
deviate from their true beliefs and misspecify their predictive
densities, with multiplication by the weight function (and subsequent
normalization) being an optimal strategy.  Therefore, the scoring rule
$\myS$ in \eqref{eq:improper} is improper.  

To connect to the forecaster's dilemma, consider the indicator weight
function $w_r(y) = \one \{ y \geq r \}$.  The use of the weight
function $w_r$ does not directly correspond to restricting the
evaluation set to cases where the observation exceeds or equals the
threshold value $r$, as instead of excluding these cases, a score of
zero is assigned to them.  However, when forecast methods are
compared, the use of the indicator weighted scoring rule corresponds
to a multiplicative scaling of the restricted score, and so the
ranking of competing forecasts is the same as that obtained by
restricting the evaluation set.

\subsection{Tailoring proper scoring rules}  \label{sec:weighted}
 
The forecaster's dilemma gives rise to the question how one might
apply scoring rules to probabilistic forecasts when particular
emphasis is placed on extreme events, while retaining propriety.  To
this end, \citet{DiksEtAl2011} and \citet{GneitingRanjan2011} consider
the use of proper weighted scoring rules that emphasize specific
regions of interest.
 
\citet{DiksEtAl2011} propose the {\em conditional likelihood}\/ (CL)
score,
\begin{equation}  \label{eq:CL}
\textnormal{CL}(F,y) = 
- w(y) \log \! \left( \frac{f(y)}{\int_{-\infty}^\infty w(z) f(z) \, \dd z} \right) \! ,
\end{equation}
and the {\em censored likelihood}\/ (CSL) score,
\begin{equation}  \label{eq:CSL}
\textnormal{CSL}(F,y) = 
- w(y) \log f(y) - (1-w(y)) \log \! \left( 1- \int_{-\infty}^\infty w(z) f(z) \, \dd z \right) \! .
\end{equation}
Here, $w$ is a weight function such that $0 \leq w(z) \leq 1$ and
$\int w(z) f(z) \, \dd z > 0$ for all potential predictive
distributions, where $f$ denotes the density of $F$.  When $w(z)
\equiv 1$, both the CL and the CSL score reduce to the unweighted
logarithmic score \eqref{eq:LogS}.  \cite{GneitingRanjan2011} propose
the {\em threshold-weighted continuous ranked probability score}\/
(twCRPS), defined as
\begin{equation}  \label{eq:twCRPS}    
\textnormal{twCRPS}(F,y) 
= \int_{-\infty}^\infty w(z) \left( F(z) - \one \{ y \leq z \} \right)^2  \, \dd z,  
\end{equation}  
where, again, $w$ is a non-negative weight function.  When $w(z)
\equiv 1$, the twCRPS reduces to the unweighted CRPS \eqref{eq:CRPS}.
For recent applications of the twCRPS and a quantile-weighted version
of the CRPS see, for example, \citet{CooleyEtAl2012},
\citet{LerchThorarinsdottir2013} and \citet{ManzanZerom2013}.

As noted, these scoring rules are proper and can be tailored to the
region of interest.  When interest centers on the right tail of the
distribution, we may choose $w(z) = \one \{ z \geq r\}$ for some high
threshold $r$.  However, the indicator weight function might result in
violations of the regularity conditions for the CL and CSL scoring
rule, unless all predictive densities considered are strictly
positive.  Furthermore, predictive distributions that are identical on
$[r,\infty)$, but differ on $(-\infty,r)$, cannot be distinguished.
Weight functions based on CDFs as proposed by
\citet{AmisanoGiacomini2007} and \citet{GneitingRanjan2011} provide
suitable alternatives.  For instance, we can set $w(z) = \Phi(z \, |
\, r, \sigma^2)$ for some $\sigma > 0$, where $\Phi( \cdot \, | \,
\mu, \sigma^2)$ denotes the CDF of a normal distribution with mean
$\mu$ and variance $\sigma^2$.  Weight functions emphasizing the left
tail of the distribution can be constructed similarly, by using $w(z)
= \one \{z \leq r \}$ or $w(z) = 1 - \Phi(z \, | \, r, \sigma^2)$ for
some low threshold $r$.  In practice, the weighted integrals in
\eqref{eq:CL}, \eqref{eq:CSL}, and \eqref{eq:twCRPS} may need to be
approximated by discrete sums, which corresponds to the use of a
discrete weight measure, rather than a weight function, as discussed
by \citet{GneitingRanjan2011}.
  
In what follows we focus on the above proper variants of the LogS and
the CRPS.  However, further types of proper weighted scoring rules can
be developed.  \citet{Pelenis2014} introduces the penalized weighted
likelihood score and the incremental CPRS.  \citet{ToedterAhrens2012}
and \citet{JuutilainenEtAl2012} propose a logarithmic scoring rule
that depends on the predictive CDF rather than the predictive density.
As hinted at by \citet[p.~466]{JuutilainenEtAl2012}, this score can be
generalized to a weighted version, which we call the {\em
  threshold-weighted continuous ranked logarithmic score}\/ (twCRLS),
\begin{equation}  \label{eq:twCRLS}
\textnormal{twCRLS}(F,y) 
= - \int_\real w(z) \log |F(z) - \one \{y > z \}| \, \dd z.
\end{equation}
In analogy to the twCRPS \eqref{eq:twCRPS} being a weighted integral
of the Brier score in \eqref{eq:BS}, the twCRLS \eqref{eq:twCRLS}
can be interpreted as a weighted integral of the discrete {\em logarithmic
score} (LS) \citep{Good1952, GneitingRaftery2007},
\begin{align}  
\textnormal{LS}_z(F,y) 
& = - \log |F(z) - \one \{y > z \}| \label{eq:LS} \\
& = - \one \{ y \leq z \} \log F(z) - \one \{ y > z \} \log \! \left( 1 - F(z) \right) \! ,  
  \nonumber
\end{align}
for the induced probability forecast for the binary event of the
observation not exceeding the threshold value $z$.  The aforementioned
weight functions and discrete approximations can be employed.

\subsection{Diebold-Mariano tests}  \label{sec:DM}
 
Formal statistical tests of equal predictive performance have been
widely used, particularly in the economic literature.  Turning now to
a time series setting, we consider probabilistic forecasts $F_t$ and
$G_t$ for an observation $y_{t+k}$ that lies $k$ time steps ahead.
Given a proper scoring rule $\myS$, we denote the respective mean
scores on a test set ranging from time $t = 1, \dots, n$ by
\[
\bar\myS^F_n = \frac{1}{n} \sum_{t=1}^n \myS(F_t,y_{t+k}) 
\qquad \textnormal{and} \qquad  
\bar\myS^G_n = \frac{1}{n} \sum_{t=1}^n \myS(G_t,y_{t+k}), 
\]
respectively.  \citet{DieboldMariano1995} proposed the use of 
the test statistic
\begin{equation}  \label{eq:DM} 
t_n = \sqrt{n} \, \frac{\bar\myS^F_n - \bar\myS^G_n}{\hat\sigma_n}, 
\end{equation}
where $\hat\sigma_n^2$ is a suitable estimator of the asymptotic
variance of the score difference.  Under the null hypothesis of a
vanishing expected score difference and standard regularity
conditions, the test statistic $t_n$ in \eqref{eq:DM} is
asymptotically standard normal \citep{DieboldMariano1995,
  GiacominiWhite2006, Diebold2012}.  When the null hypothesis is
rejected in a two-sided test, $F$ is preferred if the test statistic
$t_n$ is negative, and $G$ is preferred if $t_n$ is positive.
  
For $j = 0, 1, \ldots$ let $\hat\gamma_j$ denote the lag $j$
sample autocovariance of the sequence $\myS(F_1,y_{1+k}) -
\myS(G_1,y_{1+k}), \ldots, \myS(F_n,y_{n+k}) - \myS(G_n,y_{n+k})$ of
score differences.  \citet{DieboldMariano1995} noted that for ideal
forecasts at the $k$ step ahead prediction horizon the respective
errors are at most $(k-1)$-dependent.  Motivated by this fact,
\citet{GneitingRanjan2011} use the estimator
\begin{equation}  \label{eq:GR}
\hat\sigma_n^2 =  
\begin{cases}
\hat\gamma_0 & \textnormal{if } k = 1, \\ 
\hat\gamma_0 + 2\sum_{j=1}^{k-1} \hat\gamma_j & \textnormal{if } k \geq 2.
\end{cases}
\end{equation}
for the asymptotic variance in the test statistic \eqref{eq:DM}.
While the at most $(k-1)$-dependence assumption might be violated in
practice for various reasons, this appears to be a reasonable and
practically useful choice nonetheless.  \citet{DiksEtAl2011} propose
the use of the heteroskedasticity and autocorrelation consistent (HAC)
estimator
\begin{equation}  \label{eq:DiksEtAl} 
\hat\sigma_n^2 = \hat\gamma_0 
+ 2 \sum_{j=1}^J \! \left( 1 - \frac{j}{J} \right) \! \hat\gamma_j,
\end{equation}  
where $J$ is the largest integer less than or equal to $n^{1/4}$.
When this latter estimator is used, larger estimates of the asymptotic
variance and smaller absolute values of the test statistic
\eqref{eq:DM} tend to be obtained, as compared to using the estimator
\eqref{eq:GR}, particularly when the sample size $n$ is large.

\section{Simulation studies}  \label{sec:sim}

We now present simulation studies.  In Section \ref{sec:sim1} we mimic
the experiment reported on in Table \ref{tab:sim1a} for point
forecasts, now illustrating the forecaster's dilemma on probabilistic
forecasts.  Furthermore, we consider the influence of the
signal-to-noise ratio in the data generating process.  Thereafter in
the following Sections, we investigate whether or not there is a case
for the use of proper weighted scoring rules, as opposed to their
unweighted counterparts, when interest focuses on extremes.  As it
turns out, the fundamental lemma of \citet{NeymanPearson1933} provides
theoretical guidance in this regard.  All results in this section are
based on 10\,000 replications.

\subsection{The influence of the signal-to-noise ratio}  \label{sec:sim1}

Let us recall that in the simulation setting of eq.~\eqref{eq:sim1}
the observation satisfies $Y | \, \mu \sim \cN(\mu,\sigma^2)$ where
$\mu \sim \cN(0,1-\sigma^2)$.  In Table \ref{tab:sim1a} we have
considered three competing point forecasts --- termed the perfect,
unconditional, and extremist forecasts --- and have noted the
appearance of the forecaster's dilemma when the quality of the
forecasts is assessed on cases of extreme outcomes only.

\begin{table}[t]

\centering
 
\caption{Mean scores for the probabilistic forecasts in Table
  \ref{tab:sim1a}, where the observation $Y$ satisfies \eqref{eq:sim1}
  with $\sigma^2 = \frac{2}{3}$ being fixed.  The CRPS and LogS are
  computed based on all observations, whereas the restricted versions
  (rCRPS and rLogS) are based on observations exceeding 1.64, the 95th
  percentile of the population, only.  The lowest value in each column
  is shown in bold.
  \label{tab:sim1b}}
  
\begin{tabular}{lccccc}
\toprule
Forecast      & CRPS & LogS & rCRPS & rLogS \\
\midrule
Perfect       & \bf{0.46} & \bf{1.22} & 0.96 & 2.30 \\
Unconditional & 0.57 & 1.42  & 1.48 & 3.03 \\
Extremist     & 2.05 & 5.90 & \bf{0.79} & \bf{1.88}  \\
\bottomrule
\end{tabular}

\bigskip
 
\caption{Mean scores for the probabilistic forecasts in Table
  \ref{tab:sim1a}, where the observation $Y$ satisfies \eqref{eq:sim1}
  with $\sigma^2 = \frac{2}{3}$ being fixed, under the proper weighted
  scoring rules twCRPS, CL, and CSL.  For each weight function and
  column, the lowest value is shown in bold.  \label{tab:sim1c}}
  
\begin{tabular}{clccc}
\toprule
Threshold $r$ & Forecast & twCRPS & CL & CSL   \\
\midrule
\multicolumn{5}{l}{Indicator weight function, $w(z) = \one \{ z \geq 1.64 \}$} \\
\midrule
1.64 & Perfect       & \bf{0.018} & \bf{$<$ 0.001} & \bf{0.164} \\
     & Unconditional & 0.019 & $\hphantom{<}0.002$ & 0.204 \\
     & Extremist     & 0.575 & $\hphantom{<}0.093$ & 2.205 \\
\midrule
\multicolumn{5}{l}{Gaussian weight function, $w_r(z) = \Phi(z \, | \, 1.64, 1)$} \\
\midrule
1.64 & Perfect         & \bf{0.053} & \bf{$-$0.043} & \bf{0.298} \\
     & Unconditional & 0.062 & $-0.028$  & 0.345 \\
     & Extremist     & 0.673 & \hphantom{$-$}0.379 & 1.625 \\
\bottomrule
\end{tabular}
  
\end{table}
  
We now turn to probabilistic forecasts and study the effect of the
parameter $\sigma \in (0,1)$ that governs predictability.  Small
values of $\sigma$ correspond to high signal-to-noise ratios, and
large values of $\sigma$ to small signal-to-noise ratios,
respectively.  Marginally, $Y$ is standard normal for all values of
$\sigma$.  In the limit as $\sigma \to 0$ the perfect predictive
distribution approaches the point measure in the random mean $\mu$; as
$\sigma \to 1$ it approaches the unconditional standard normal
distribution.  The perfect probabilistic forecast is ideal in the
technical sense of Section \ref{sec:framework} and thus will be
preferred over any other predictive distribution (with identical
information basis) by any rational user \citep{DieboldEtAl1998,
Tsyplakov2013}.

In Table \ref{tab:sim1b} we report mean scores for the three
probabilistic forecasts when $\sigma^2 = \frac{2}{3}$ is fixed.  Under
the CRPS and LogS the perfect forecast outperforms the others, as
expected, and the extremist forecast performs by far the worst.
However, these results change drastically if cases with extreme
observations are considered only.  In analogy to the results in Table
\ref{tab:sim1a}, the perfect forecast is discredited under the
restricted scores rCRPS and rLogS, whereas the misguided extremist
forecast appears to excel, thereby demonstrating the forecaster's
dilemma in the setting of probabilistic forecasts.  As shown in Table
\ref{tab:sim1c}, under the proper weighted scoring rules introduced in
Section \ref{sec:weighted} with weight functions that emphasize the
right tail, the rankings under the unweighted CRPS and LogS are
restored.

\begin{figure}[t]

\centering
 
\includegraphics[width=\textwidth]{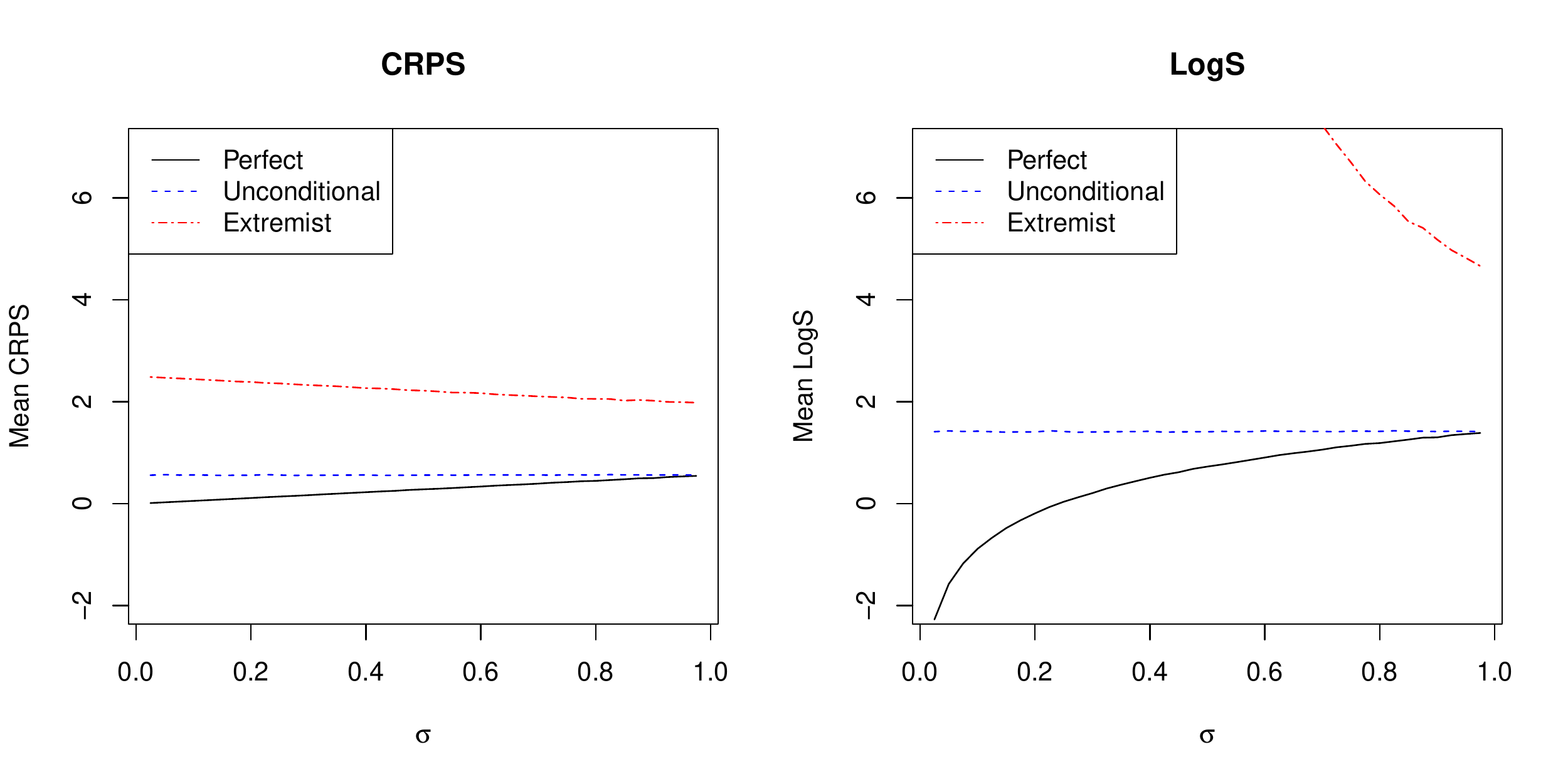}

\vspace{-4mm} 

\caption{Mean CRPS and LogS for the probabilistic forecasts in the
  setting of eq.~\eqref{eq:sim1} and Table \ref{tab:sim1a} as
  functions of the parameter $\sigma \in (0,1)$. \label{fig:sim1b}}

\smallskip

\includegraphics[width=\textwidth]{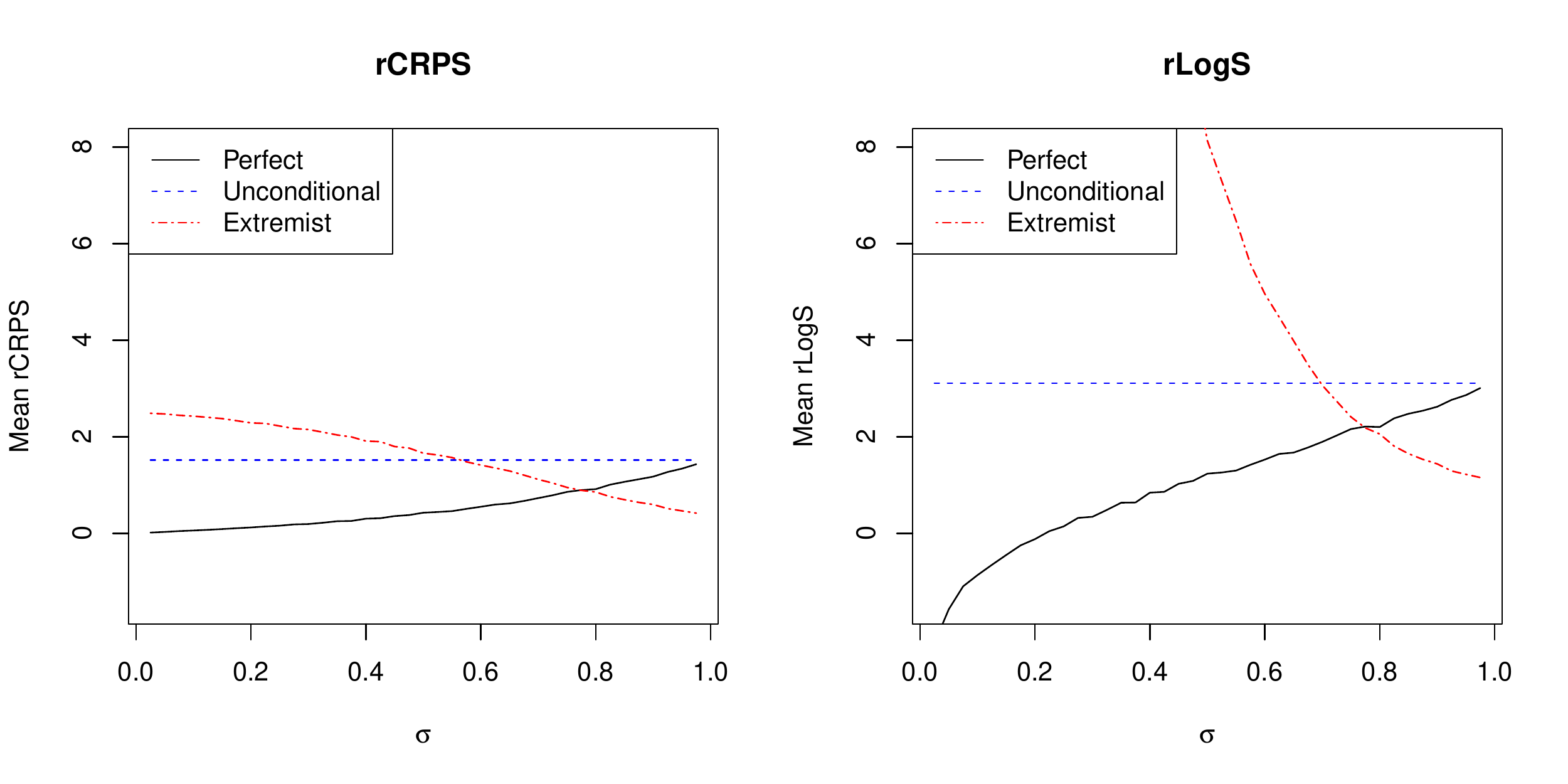}

\vspace{-4mm} 
 
\caption{Mean of the improper restricted scoring rules rCRPS and rLogS
  for the probabilistic forecasts in the setting of
  eq.~\eqref{eq:sim1} and Table \ref{tab:sim1a} as functions of the
  parameter $\sigma \in (0,1)$.  The restricted mean scores are based
  on the subset of observations exceeding 1.64 only.
  \label{fig:sim1c}}

\end{figure}
 
Next we investigate the influence of the signal-to-noise ratio in the
data generating process on the appearance and extent of the
forecaster's dilemma.  As noted, predictability increases with the
parameter $\sigma \in (0,1)$.  Figure \ref{fig:sim1b} shows the mean
CRPS and LogS for the three probabilistic forecasts as a function of
$\sigma$.  The scores for the unconditional forecast do not depend on
$\sigma$.  The predictive performance of the perfect forecast
decreases in $\sigma$, which is natural, as it is less beneficial to
know the value of $\mu$ when $\sigma$ is large.  The extremist
forecast yields better scores as $\sigma$ increases, which can be
explained by the increase in the predictive variance that allows for a
better match between the probabilistic forecast and the true
distribution.  For the improper restricted scoring rules rCRPS and
rLogS, the same general patterns can be observed in Figure
\ref{fig:sim1c} --- the mean score increases in $\sigma$ for the
perfect forecast and decreases for the extremist forecast.  In
accordance with the forecaster's dilemma, the extremist forecast is
now perceived to outperform its competitors for all sufficiently large
values of $\sigma$.  However, for small values of $\sigma$, when the
signal in $\mu$ is strong, the rankings are the same as under the CRPS
and LogS in Figure \ref{fig:sim1b}.  This illustrates the intuitively
obvious observation that the forecaster's dilemma is tied to
stochastic systems with moderate to low signal-to-noise ratios, so
that predictability is weak.

\subsection{Power of Diebold-Mariano tests: \citet{DiksEtAl2011} revisited}  \label{sec:sim2a}

While thus far we have illustrated the forecaster's dilemma, the
unweighted CRPS and LogS are well able to distinguish between the
perfect forecast and its competitors.  In the subsequent sections we
investigate whether there are benefits to using proper weighted
scoring rules, as opposed to their unweighted versions.

To begin with, we adopt the simulation setting in Section 4 of
\citet{DiksEtAl2011}.  Suppose that at time $t = 1, \ldots, n$, the
observations $y_t$ are independent standard normal.  We apply the
two-sided Diebold-Mariano test of equal predictive performance to
compare the ideal probabilistic forecast, the standard normal
distribution, to a misspecified competitor, a Student $t$ distribution
with five degrees of freedom, mean 0, and variance 1.  Following
\citet{DiksEtAl2011}, we use the nominal level $0.05$, the variance
estimate \eqref{eq:DiksEtAl}, and the indicator weight function $w(z)
= \one \{z \leq r \}$, and we vary the sample size, $n$, with the
threshold value $r$ in such a way that under the standard normal
distribution the expected number, $c = 5$, of observations in the
relevant region $(-\infty,r]$ remains constant.

\begin{figure}[t]
 
\centering

\includegraphics[width=0.95\textwidth]{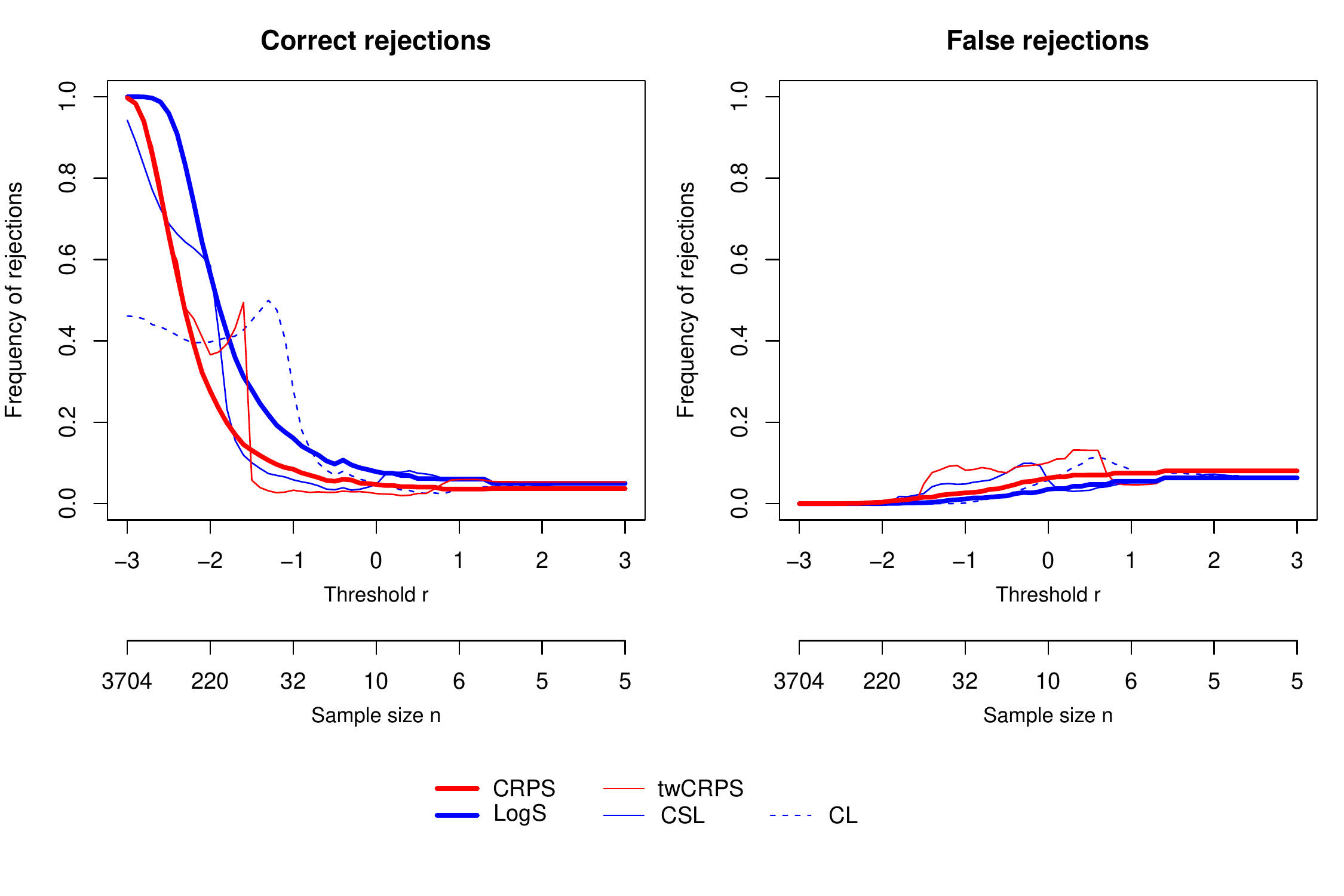}
\vspace{-5mm}

\caption{Frequency of correct rejections (in favor of the standard
  normal distribution, left panel) and false rejections (in favor of
  the Student $t$ distribution, right panel) in two-sided
  Diebold-Mariano tests in the simulation setting described in Section
  \ref{sec:sim2a}.  The panels correspond to those in the left hand
  column of Figure 5 in \citet{DiksEtAl2011}.  The sample size $n$ for
  the tests depends on the threshold $r$ in the indicator weight
  function $w(z) = \one \{z \leq r\}$ for the twCRPS, CL, and CSL
  scoring rules such that under the standard normal distribution there
  are five expected observations in the relevant interval
  $(-\infty,r]$.  \label{fig:DiksEtAl.Phi}}

\end{figure}

Figure \ref{fig:DiksEtAl.Phi} shows the proportion of rejections of
the null hypothesis of equal predictive performance in favor of either
the standard normal or the Student $t$ distribution, respectively, as
a function of the threshold value $r$ in the weight function.
Rejections in favor of the standard normal distribution represent true
power, whereas rejections in favor of the misspecified Student $t$
distribution are misguided.  The curves for the tests based on the
twCRPS, CL, and CSL scoring rules agree with those in the left column
of Figure 5 of \cite{DiksEtAl2011}.  At first sight, they might
suggest that the use of the indicator weight function $w(z) = \one \{z
\leq r \}$ with emphasis on the extreme left tail, as reflected by
increasingly smaller values of $r$, yields increased power.  At second
sight, we need to compare to the power curves for tests using the
unweighted CRPS and LogS, based on the same sample size, $n$, as
corresponds to the threshold $r$ at hand.  These curves suggest,
perhaps surprisingly, that there may not be not be an advantage to
using weighted scoring rules.  To the contrary, the left-hand panel in
Figure \ref{fig:DiksEtAl.Phi} suggests that tests based on the
unweighted LogS are competitive in terms of statistical power.

\subsection{The role of the Neyman-Pearson lemma}  \label{sec:NP}

In order to understand this phenomenon, we follow the lead of
\citet{FeuervergerRahman1992} and draw a connection to a cornerstone
of test theory, namely, the fundamental lemma of
\citet{NeymanPearson1933}.  In doing so we consider, for the moment,
one-sided rather than two-sided tests.

In the simulation setting described by \citet{DiksEtAl2011} and in the
previous section, any test of equal predictive performance can be
re-interpreted as a test of the simple null hypothesis $H_0$ of a
standard normal population against the simple alternative $H_1$ of a
Student $t$ population.  We write $f_0$ and $f_1$ for the
associated density functions and $\myP_0$ and $\myP_1$ for probabilities
under the respective hypotheses.  By the Neyman-Pearson lemma
\citep[Theorem 3.2.1]{LehmannRomano2005}, under $H_0$ and at any level
$\alpha \in (0,1)$ the unique most powerful test of $H_0$ against
$H_1$ is the likelihood ratio test.  The likelihood ratio test rejects
$H_0$ if $\prod_{t = 1}^n f_1(y_t) / \prod_{t = 1}^n f_0(y_t) > k$ or,
equivalently, if
\begin{equation}  \label{eq:LRT}  
\sum_{t = 1}^n \log f_1(y_t) - \sum_{t = 1}^n \log f_0(y_t) > \log k,   
\end{equation} 
where the critical value $k$ is such that 
\[
\myP_0 \! \left( \frac{\prod_{t = 1}^n f_1(y_t)}{\prod_{t = 1}^n f_0(y_t)} > k \right) 
= \alpha.  
\]
Due to the optimality property of the likelihood ratio test, its
power, 
\begin{equation}  \label{eq:beta}  
\myP_1 \! \left( \frac{\prod_{t = 1}^n f_1(y_t)}{\prod_{t = 1}^n f_0(y_t)} > k \right) \! , 
\end{equation}
gives a theoretical upper bound on the power of any test of $H_0$
versus $H_1$.  Furthermore, the optimality result is robust, in the
technical sense that minor misspecifications of either $H_0$ or $H_1$,
as quantified by the Kullback-Leibler divergence, lead to minor loss of
power only \citep{EguchiCopas2006}.

We now compare to the one-sided Diebold-Mariano test based on the
logarithmic score (LogS; eq.~\ref{eq:LogS}).  This test uses the
statistic \eqref{eq:DM} and rejects $H_0$ if
\begin{equation}  \label{eq:DM.LRT}  
\sum_{t = 1}^n \log f_1(y_t) - \sum_{t = 1}^n \log f_0(y_t) 
> \sqrt{n} \, \hat\sigma_n z_{1 - \alpha},     
\end{equation} 
where $z_{1 - \alpha}$ is a standard normal quantile and
$\hat\sigma_n^2$ is given by \eqref{eq:GR} or \eqref{eq:DiksEtAl}.
Comparing with \eqref{eq:LRT}, we see that the one-sided
Diebold-Mariano test that is based on the LogS has the same type of
rejection region as the likelihood ratio test.  However, the
Diebold-Mariano test uses an estimated critical value, which may lead
to a level less or greater than the nominal level, $\alpha$, whereas
the likelihood ratio test uses the (in the practice of forecasting
unavailable) critical value that guarantees the desired nominal level,
$\alpha$.

\begin{figure}[t]
 
\centering
 
\includegraphics[width=0.95\textwidth]{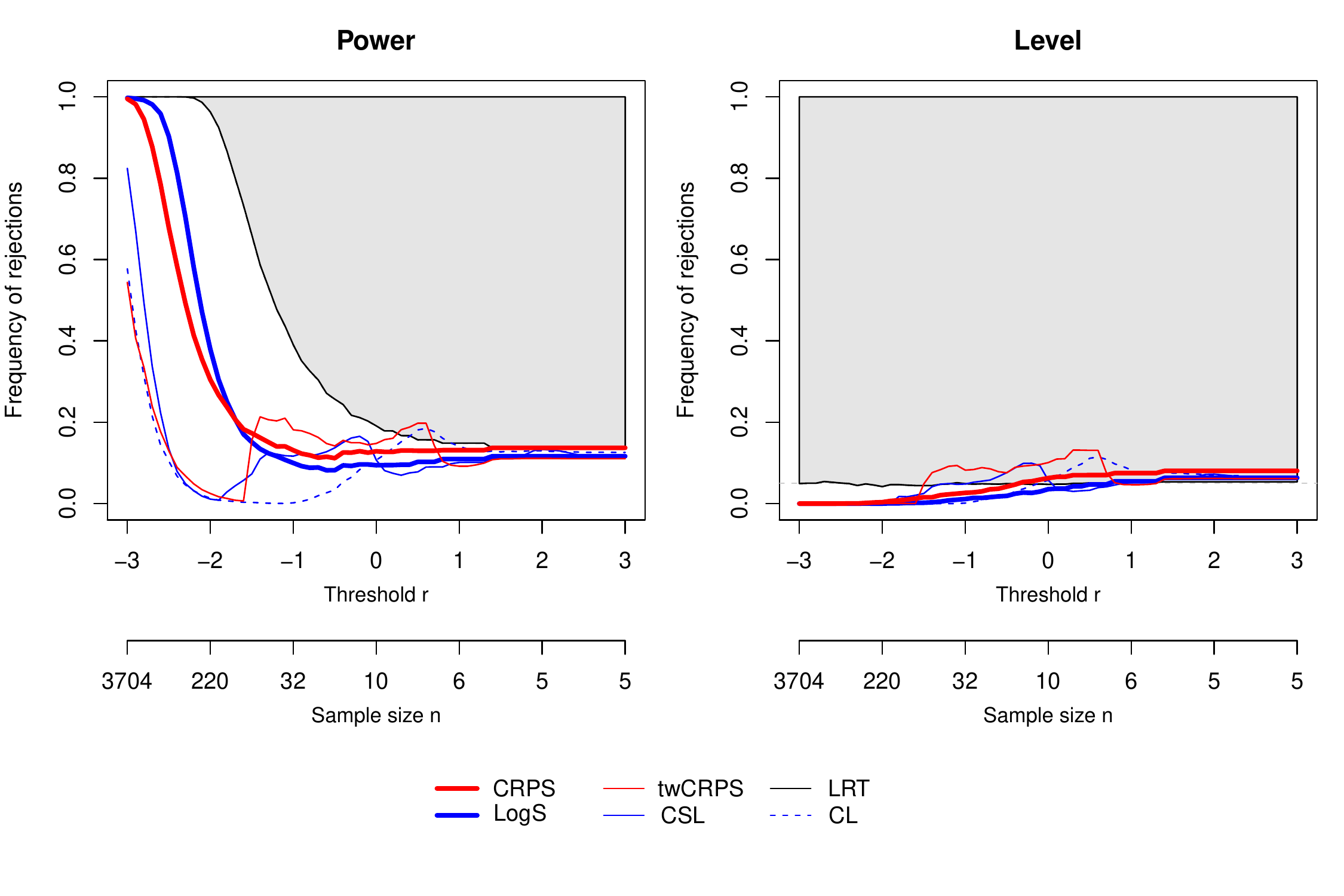}
\vspace{-5mm}

\caption{Power (left) and level (right) of the likelihood ratio test (LRT)
  and one-sided Diebold-Mariano tests in the simulation setting
  described in Section \ref{sec:sim2a}.  The sample size $n$ for the
  tests depends on the threshold $r$ in the indicator weight function
  $w(z) = \one \{z \leq r\}$ for the twCRPS, CL, and CSL scoring rules
  such that under the standard normal distribution there are five
  expected observations in the relevant interval $(-\infty,r]$.  In
  the panel for power, the shaded area above the curve for the
  LRT corresponds to theoretically unattainable
  values for a test with nominal level.  In the panel for level, the
  dashed line indicates the nominal level.  \label{fig:NP}}

\end{figure}

In this light, it is not surprising that the one-sided Diebold-Mariano
test based on the LogS has power close to the theoretical optimum in
\eqref{eq:beta}.  We illustrate this in Figure \ref{fig:NP}, where we
plot the power and size of the likelihood ratio test and one-sided
Diebold-Mariano tests based on the CRPS, twCRPS, LogS, CL, and CSL in
the setting of the previous section.  For small threshold values, the
Diebold-Mariano test based on the unweighted LogS has much higher
power than tests based on the weighted scores, even though it does not
reach the power of the likelihood ratio test, which can be explained
by the use of an estimated critical value and incorrect size
properties.  The theoretical upper bound on the power is violated by
Diebold-Mariano tests based on the twCRPS and CL for threshold values
between 0 and 1.  However, the level of these tests exceeds the
nominal level of $\alpha = 0.05$ with too frequent rejections of
$H_0$.

In the setting of two-sided tests, the connection to the
Neyman-Pearson lemma is less straightforward, but the general
principles remain valid and provide a partial explanation of the
behavior seen in Figure \ref{fig:DiksEtAl.Phi}.

\subsection{Power of Diebold-Mariano tests: Further experiments}  \label{sec:sim2b}

In the simulation experiments just reported, Diebold-Mariano tests
based on proper weighted scoring rules generally are unable to
outperform tests based on traditionally used, unweighted scoring
rules.  Several potential reasons come to mind.  As we have just seen,
when the true data generating process is given by one of the competing
forecast distributions, the Neyman-Pearson lemma points at the
superiority of tests based on the unweighted LogS.  Furthermore, in
the simulation setting considered thus far, the distributions
considered differ both in the center, the left tail, and the right
tail, and the test sample size varied with the threshold for the
weight function in a peculiar way.  

Therefore, we now consider a revised simulation setting, where we
compare two forecast distributions neither of which corresponds to the
true sampling distribution, where the forecast distributions only
differ on the positive half-axis, and where the test sample size is
fixed at $n = 100$.  The three candidate distributions are given by
$\Phi$, a standard normal distribution with density $\phi$, by a
heavy-tailed distribution $H$ with density
\[ 
h(x) = \one \{x \leq 0 \} \, \phi(x) + \one\{x > 0\} \, \frac{3}{8}\left( 1+ \frac{x^2}{4} \right)^{-5/2}, 
\] 
and by an equally weighted
mixture $F$ of $\Phi$ and $H$, with density
\[ 
f(x) = \frac{1}{2} \left( \phi(x) + h(x) \right) \! .  
\] 
We perform two-sided Diebold-Mariano tests of equal predictive
performance based on the CRPS, twCRPS, LogS, CL, and CSL.  

In Scenario A, the data are a sample from the standard normal
distribution $\Phi$, and we compare the forecasts $F$ and $H$,
respectively.  In Scenario B, we interchange the roles of $\Phi$ and
$H$, that is, the data are a sample from $H$, and we compare the
forecasts $F$ and $\Phi$.  The Neyman-Pearson lemma does not apply in
this setting.  However, the definition of $F$ as a weighted mixture of
the true distribution and a misspecified competitor lets us expect
that $F$ is to be preferred over the latter.  Indeed, by Proposition 3
of \citet{Nau1985}, if $F = w \,G + (1-w) \, H$ with $w\in [0,1]$ is a
convex combination of $G$ and $H$, then
\[
\myE_G \, \myS(G,Y) \leq \myE_G \, \myS(F,Y) \leq \myE_G \, \myS(H,Y)
\]
for any proper scoring rule $\myS$.  As any utility function induces a
proper scoring rule via the respective Bayes act, this implies that
under $G$ any rational decision maker favors $F$ over $H$
\citep{Dawid2007, GneitingRaftery2007}.

We estimate the frequencies of rejections of the null hypothesis of
equal predictive performance at level $\alpha = 0.05$.  The choice of
the estimator for the asymptotic variance of the score difference in
the Diebold-Mariano test statistic \eqref{eq:DM} does not have a
recognizable effect in this setting, and so we show results under the
estimator \eqref{eq:GR} with $k = 1$ only.

\begin{figure}[p] 

\centering 

\includegraphics[width = \textwidth]{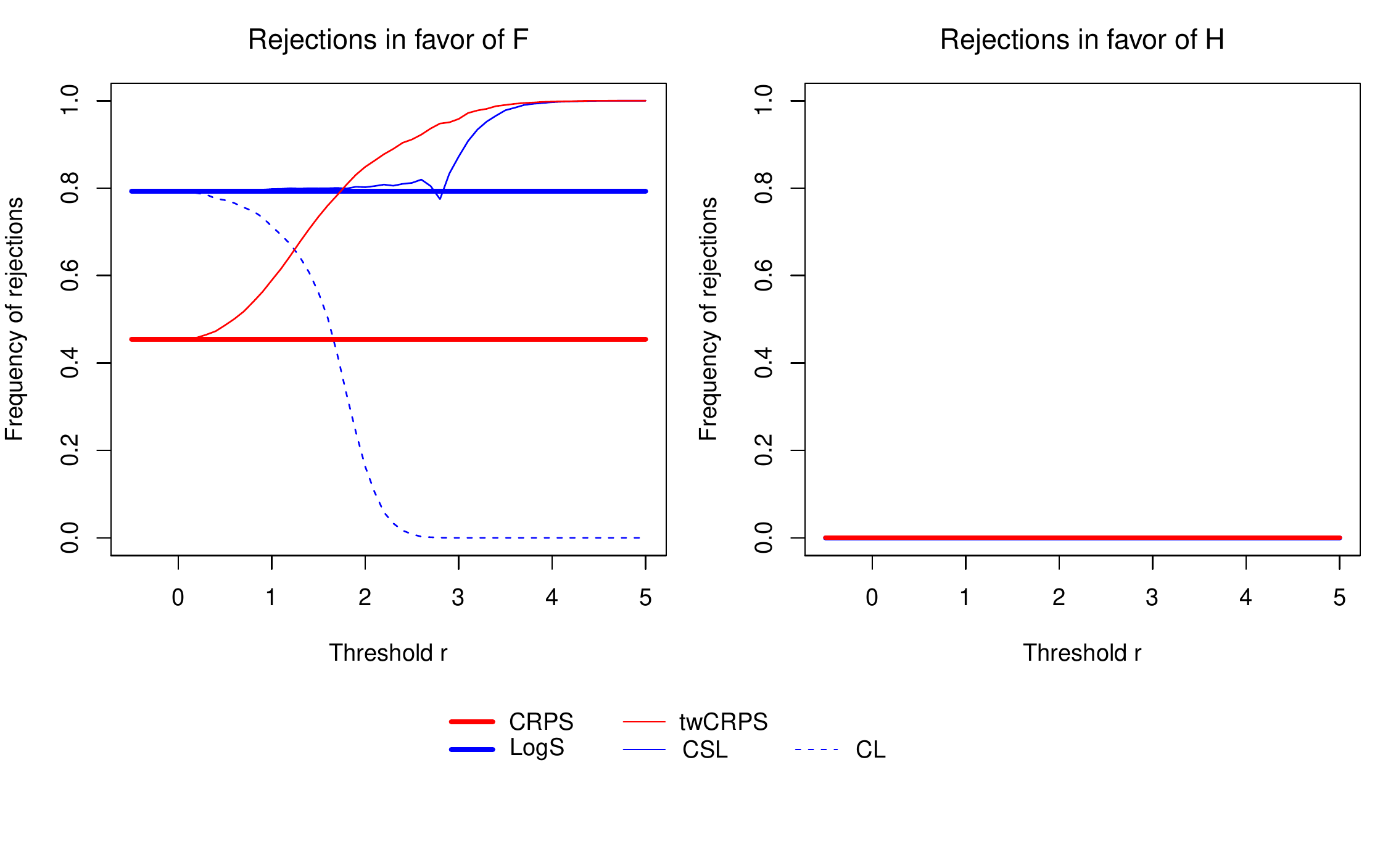} 
\vspace{-12mm}

\caption{Scenario A in Section
  \ref{sec:sim2b}.  The null hypothesis of equal predictive
  performance of $F$ and $H$ is tested under a standard normal
  population.  The panels show the frequency of rejections in
  two-sided Diebold-Mariano tests in favor of either $F$ (left,
  desired) or $H$ (misguided, right).  The tests under the twCRPS, CL,
  and CSL scoring rules use the weight function $w(z) = \one \{z \geq
  r \}$, and the sample size is fixed at $n = 100$.  \label{fig:Phi}}

\bigskip
\bigskip

\includegraphics[width = \textwidth]{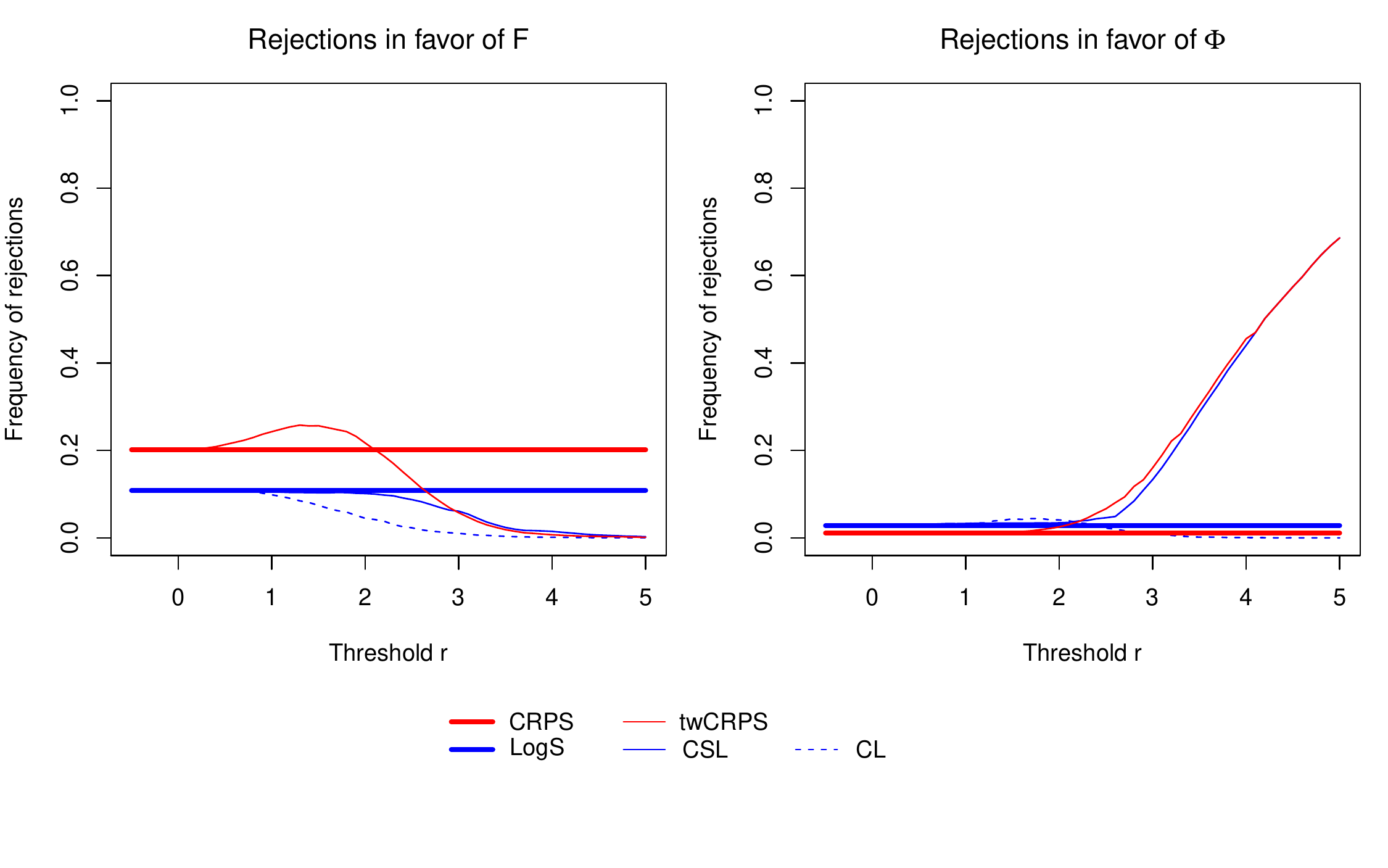} 
\vspace{-12mm}

\caption{Scenario B in Section
  \ref{sec:sim2b}.  The null hypothesis of equal predictive
  performance of $F$ and $\Phi$ is tested under a Student $t$
  population.  The panels show the frequency of rejections in
  two-sided Diebold-Mariano tests in favor of either $F$ (desired,
  left) or $\Phi$ (misguided, right).  The tests under the twCRPS, CL,
  and CSL scoring rules use the weight function $w(z) = \one \{z \geq
  r \}$, and the sample size is fixed at $n = 100$.  \label{fig:t}}

\end{figure}

Figure \ref{fig:Phi} shows rejection rates under Scenario A in favor
of $F$ and $H$, respectively, as a function of the threshold $r$ in
the indicator weight function $w(z) = \one \{z \geq r \}$ for the
weighted scoring rules.  The frequency of the desired rejections in
favor of $F$ increases with larger thresholds for tests based on the
twCRPS and CSL, thereby suggesting an improved discrimination ability
at high threshold values.  Under the CL scoring rule, the rejection
rate decreases rapidly for larger threshold values.  This can be
explained by the fact that the weight function is a multiplicative
component of the CL score in \eqref{eq:CL}.  As $r$ becomes larger and
larger, none of the 100 observations in the test sample exceed the
threshold, and so the mean scores under both forecasts vanish.  This
can also be observed in Figure \ref{fig:DiksEtAl.Phi}, where, however,
the effect is partially concealed by the increase of the sample size
for more extreme threshold values.  Interestingly, an issue very
similar to that for the CL scoring rule arises in the assessment of
deterministic forecasts of rare and extreme binary events, where
performance measures based on contingency tables have been developed
and standard measures degenerate to trivial values as events become
rarer \citep{Marzban1998, StephensonEtAl2008}, posing a challenge that
has been addressed by \citet{FerroStephenson2011}.

Figure \ref{fig:t} shows the respective rejection rates under Scenario
B, where the sample is generated from the heavy-tailed distribution
$H$, and the forecasts $F$ and $\Phi$ are compared.  In contrast to
the previous examples the Diebold-Mariano test based on the CRPS shows
a higher frequency of the desired rejections in favor of $F$ than the
test based on the LogS.  However, for the tests based on proper
weighted scoring rules, the frequency of the desired rejections in
favor of $F$ decays to zero with increasing threshold value, and for
the tests based on the twCRPS and CSL, the frequency of the undesired
rejections in favor of $\Phi$ rises for larger threshold values.

This seemingly counterintuitive observation can be explained by the
tail behavior of the forecast distributions, as follows.  Consider the
twCRPS and CSL with the indicator weight function $w(z) = \one \{z
\geq r \}$ and a threshold $r$ that exceeds the maximum of the given
sample.  In this case, the scores do not depend on the observations,
and are solely determined by the respective tail probabilities, with
the lighter tailed forecast distribution receiving the better score.
In a nutshell, when the emphasis lies on a low-probability region with
few or no observations, the forecaster assigning smaller probability
to this region will be preferred.  The traditionally used unweighted
scoring rules do not depend on a threshold and thus do not suffer from
this deficiency.

In comparisons of the mixture distribution $F$ and the lighter-tailed
forecast distribution $\Phi$ this leads to a loss of finite sample
discrimination ability of the proper weighted scoring rules as the
threshold $r$ increases.  This observation also suggests that any
favorable finite sample behavior of the Diebold-Mariano tests based on
weighted scoring rules in Scenario A might be governed by rejections
due to the lighter tails of $F$ compared to $H$.

In summary, even though the simulation setting at hand was
specifically tailored to benefit proper weighted scoring rules, these
do not consistently perform better in terms of statistical power when
compared to their unweighted counterparts.  Any advantages vanish at
increasingly extreme threshold values in case the actually superior
distribution has heavier tails.

\section{Case study}  \label{sec:cs}
 
Based on the work of \citet{ClarkRavazzolo2014}, we compare
probabilistic forecasting models for key macroeconomic variables for
the United States, serving to demonstrate the forecaster's dilemma and
the use of proper weighted scoring rules in an application setting.

\subsection{Data}  \label{sec:data}
 
\begin{figure}[t]
  
\centering
\includegraphics[width=\textwidth]{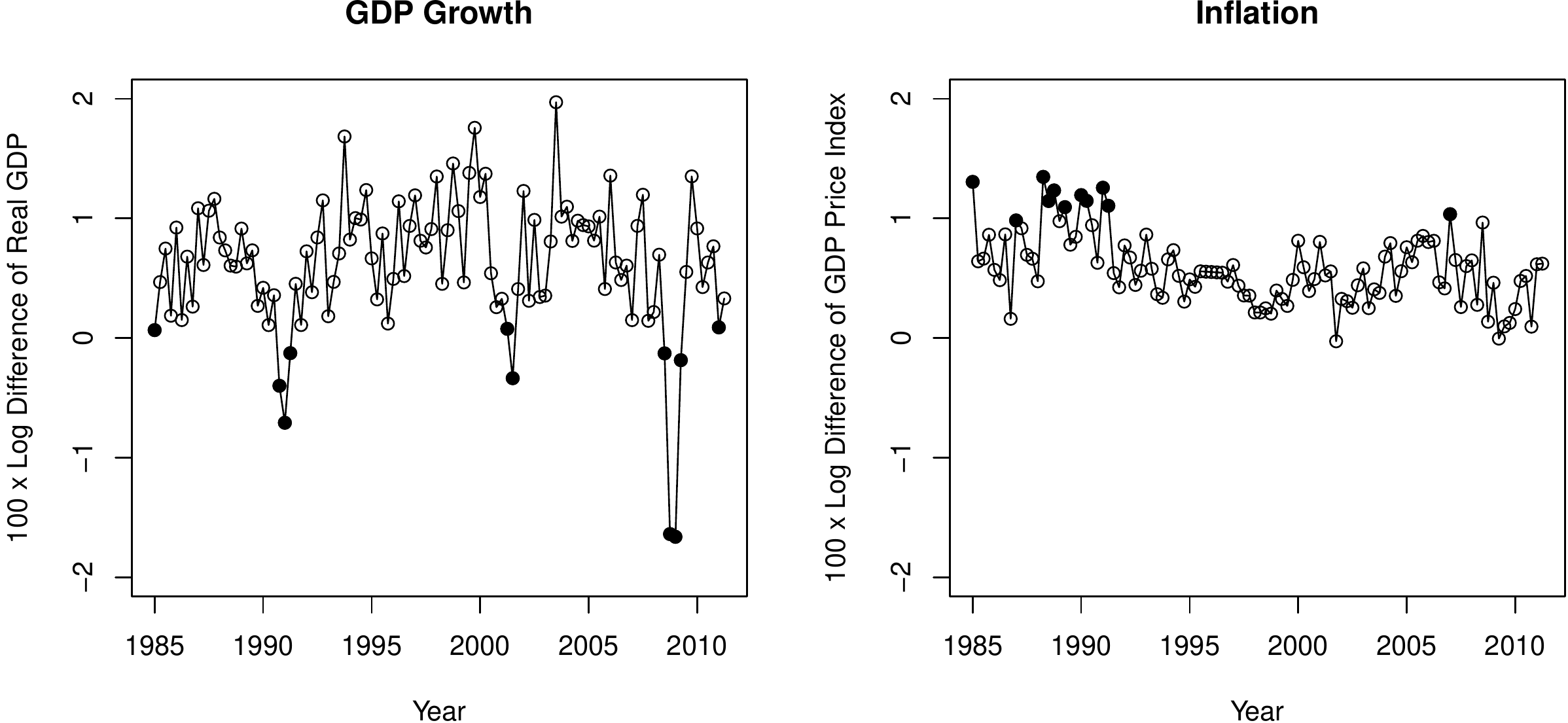}

\caption{Observations of GDP growth and inflation in the U.S.~from the
  first quarter of 1985 to the second quarter of 2011.  Solid circles
  indicate observations considered here as extreme
  events. \label{fig:data}}
 
\end{figure}

We consider time series of quarterly gross domestic product (GDP)
growth, computed as 100 times the log difference of real GDP, and
inflation in the GDP price index (henceforth {\em inflation}),
computed as 100 times the log difference of the GDP price index, over
an evaluation period from the first quarter of 1985 to the second
quarter of 2011, as illustrated in Figure \ref{fig:data}.  The data
are available from the Federal Reserve Bank of Philadelphia's real
time
dataset.\footnote{\url{http://www.phil.frb.org/research-and-data/real-time-center/real-time-data/}}

For each quarter $t$ in the evaluation period, we use the real-time
data vintage $t$ to estimate the forecasting models and construct
forecasts for period $t$ and beyond.  The data vintage $t$ includes
information up to time $t-1$.  The one-quarter ahead forecast is thus
a current quarter ($t$) forecast, while the two-quarter ahead forecast
is a next quarter ($t+1$) forecast, and so forth
\citep{ClarkRavazzolo2014}.  Here we focus on forecast horizons of one
and four quarters ahead.
 
As the GDP data are continually revised, it is not immediate which
revision should be used as the realized observation.  We follow
\citet{RomerRomer2000} and \citet{FaustWright2009} who use the second
available estimates as the actual data.  Specifically, suppose a
forecast for quarter $t + k$ is issued based on the vintage $t$ data
ending in quarter $t - 1$.  The corresponding realized observation is
then taken from the vintage $t + k + 2$ data set.  This approach may
entail structural breaks in case of benchmark revisions, but is
comparable to real-world forecasting situations where noisy early
vintages are used to estimate predictive models
\citep{FaustWright2009}.

\subsection{Forecasting models}  \label{sec:models} 

We consider autoregressive (AR) and vector autoregressive (VAR)
models, the specifications of which are given now.  For further
details and a discussion of alternative models, see
\citet{ClarkRavazzolo2014}.

Our baseline model is an AR($p$) scheme with constant shock variance.
Under this model, the conditional distribution of $Y_t$ is given by
\begin{equation}  \label{eq:AR}
Y_t \, | \, \by_{<t}, b_0, \ldots, b_p, \sigma \sim 
\cN \left( b_0 + \sum_{i=1}^p b_i y_{t-i}, \sigma^2 \right) \! ,
\end{equation}
where $p = 2$ for GDP growth and $p = 4$ for inflation.  Here,
$\by_{<t}$ denotes the vector of the realized values of the variable
$Y$ prior to time $t$.  We estimate the model parameters $b_0, \ldots,
b_p$ and $\sigma$ in a Bayesian fashion using Markov chain Monte Carlo
(MCMC) under a recursive estimation scheme, where the data sample
$\by_{<t}$ is expanded as forecasting moves forward in time.  The
predictive distribution then is the Gaussian variance-mean mixture
\begin{equation}  \label{eq:pred.distr} 
\frac{1}{m} \sum_{j=1}^m 
\cN \! \left( b_{0,j} + \sum_{i=1}^p b_{i,j} y_{t-i}, \, \sigma_j^2 \right) \! ,
\end{equation} 
where $m = 5\,000$ and $(b_{0,1}, \ldots, b_{p,1}, \sigma_1), \ldots,
(b_{0,m}, \ldots, b_{p,m}, \sigma_m)$ is a sample from the posterior
distribution of the model parameters.  For the other forecasting
models, we proceed analogously.

A more flexible approach is the Bayesian AR model with time-varying
parameters and stochastic specification of the volatility (AR-TVP-SV)
proposed by \citet{CogleySargent2005}, which has the hierarchical
structure given by
\begin{align}  \label{eq:AR-TVP-SV}
Y_t \, | \, \by_{<t}, b_{0,t}, \ldots, b_{p,t}, \lambda_t & 
\sim \cN \! \left( b_{0,t} + \sum_{i=1}^p b_{i,t} y_{t-i}, \, \lambda_t \right) \! , \\
b_{i,t} \, | \, b_{i,t-1}, \tau & \sim \cN \! \left( b_{i,t-1}, \tau^2 \right) \! , \quad i = 0, \ldots, p, \nonumber \\
\log \lambda_t \, | \, \lambda_{t-1}, \sigma & \sim \cN \! \left( \log \lambda_{t-1}, \sigma^2 \right) \! .
\nonumber
\end{align}
Again, we set $p = 2$ for GDP growth and $p = 4$ for inflation.
 
In a multivariate extension of the AR models, we consider VAR schemes
where GDP growth, inflation, unemployment rate, and three-month
government bill rate are modeled jointly.  Specifically, the
conditional distribution of the four-dimensional vector $\bY_t$ is
given by the multivariate normal distribution
\begin{equation}  \label{eq:VAR}
\bY_t \, | \, \bY_{<t}, \bb_0, \bB_1, \ldots, \bB_p, \bSigma 
\sim \cN_4 \! \left( \bb_0 + \sum_{i=1}^p \bB_i \by_{t-1}, \bSigma \right) \! ,
\end{equation}
where $\bY_{<t}$ denotes the data prior to time $t$, $\bSigma$ is a $4
\times 4$ covariance matrix, $\bb_0$ is a vector of intercepts, and
$\bB_i$ is a $4 \times 4$ matrix of lag $i$ coefficients, where $i =
1, \ldots, p$.  Here we take $p = 4$.  The univariate predictive
distributions for GDP growth and inflation arise as the respective
margins of the multivariate posterior predictive distribution.

Finally, we consider a VAR model with time-varying parameters and
stochastic specification of the volatility (VAR-TVP-SV), which is a
multivariate extension of the AR-TVP-SV model
\citep{CogleySargent2005}.  Let $\bbeta_t$ denote the vector of size
$4(4p + 1)$ comprising the parameters $\bb_{0,t}$ and $\bB_{1,t}, \ldots,
\bB_{p,t}$ at time $t$, set $\bLambda_t = \textup{diag}(\lambda_{1,t},
\ldots, \lambda_{4,t})$ and let $\bA$ be a lower triangular matrix
with ones on the diagonal and non-zero random coefficients below the
diagonal.  The VAR-TVP-SV model takes the hierarchical form
\begin{align}  \label{eq:VAR-TVP-SV}
\bY_t \, | \, \bY_{<t}, \bbeta_t, \bLambda_t, \bA & 
\sim \cN_4 \! \left( \bb_{0,t} + \sum_{i=1}^p \bB_{i,t} \by_{t-1}, \bA^{-1} \bLambda_t (\bA^{-1})^\top \right) \! , \\
\bbeta_t \, | \, \bbeta_{t-1}, \bQ & \sim \cN_{4(4p+1)} \! \left( \bbeta_{t-1}, \bQ \right) \! , \nonumber \\
\log \lambda_{i,t} \, | \, \lambda_{i,t-1}, \sigma_i & 
\sim \cN \! \left( \log \lambda_{i,t-1}, \sigma_i^2 \right) \! , \quad i = 1, \ldots, 4. \nonumber 
\end{align}
We set $p = 2$ and refer to \citet{ClarkRavazzolo2014} for further
details of the notation, the model, and its estimation.

\begin{figure}[t]
  
\centering
   
\includegraphics[width=\textwidth]{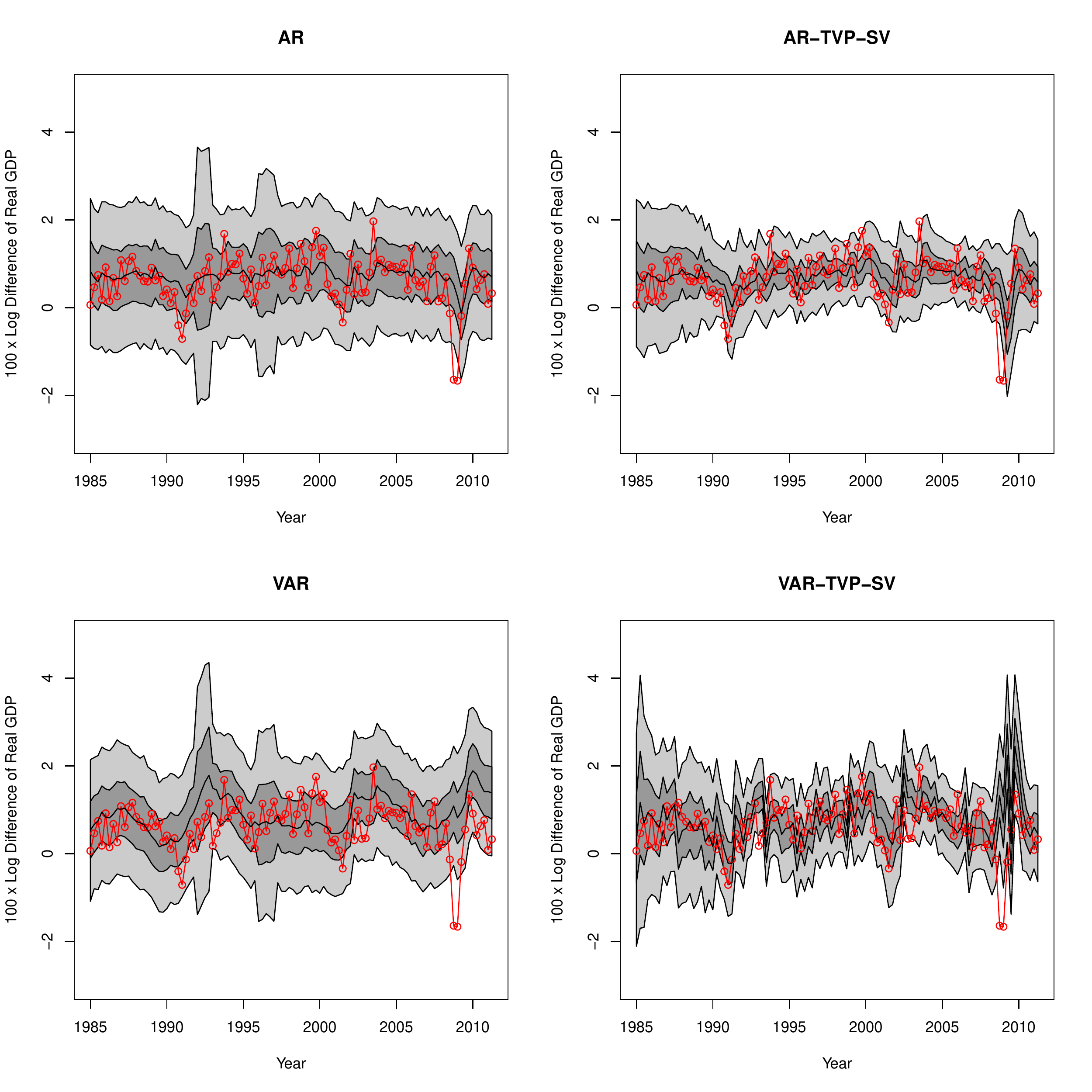}
   
\caption{One-quarter ahead forecasts of U.S.~GDP growth generated by
  the AR, AR-TVP-SV, VAR, and VAR-TVP-SV models.  The median of the
  predictive distribution is shown in the black solid line, and the
  central 50\% and 90\% prediction intervals are shaded in dark and
  light gray, respectively.  The red line shows the
  corresponding observations.  \label{fig:forecasts}}
  
\end{figure}

Figure \ref{fig:forecasts} shows one-quarter ahead forecasts of GDP
growth over the evaluation period.  The baseline models with constant
volatility generally exhibit wider prediction intervals, while the
TVP-SV models show more pronounced fluctuations both in the median
forecast and the associated uncertainty.  In 1992 and 1996, the Bureau
of Economic Analysis performed benchmark data revisions, which causes
the prediction uncertainty of the baseline models to increase
substantially.  The more flexible TVP-SV models seem less sensitive to
these revisions.

\begin{table}[t]
   
\caption{Mean CRPS and mean LogS for probabilistic forecasts of GDP
  growth and inflation in the U.S.~at prediction horizons of $k = 1$
  and $k = 4$ quarters, respectively, for the first quarter of 1985 to
  the second quarter of 2011.  For each variable and column, the
  lowest value is in bold.  \label{tab:cs1}}
   
\centering
   
\begin{tabular}{l@{\hskip 1.25cm}cc@{\hskip 1.25cm}cc}
\toprule
& \multicolumn{2}{c}{\hspace{-1.25cm}CRPS} & \multicolumn{2}{c}{LogS} \\
\midrule
& $k = 1$ & $k = 4$ & $k = 1$ & $k = 4$ \\
\midrule
GDP Growth \\
\midrule
AR	   & 0.330          & 0.359          & 1.044	      & 1.120 \\
AR-TVP-SV  & \bf{0.292} & \bf{0.329} & \bf{0.833} & \bf{1.019} \\
VAR	   & 0.385          & 0.402          & 1.118	      & 1.163 \\
VAR-TVP-SV & 0.359          & 0.420          & 0.997	      & 1.257 \\
\midrule
Inflation \\
\midrule
AR	   & 0.167	    & 0.187	     & 0.224	      & 0.374 \\
AR-TVP-SV  & \bf{0.143} & \bf{0.156} & \bf{0.047} & \bf{0.175} \\
VAR	   & 0.170	    & 0.198	     & 0.235	      & 0.428 \\
VAR-TVP-SV & 0.162	    & 0.201	     & 0.179	      & 0.552 \\
\bottomrule
\end{tabular}
   
\end{table}

\begin{table}[t]

\caption{Mean restricted CRPS (rCRPS) and restricted LogS (rLogS) for
  probabilistic forecasts of GDP growth and inflation in the U.S.~at
  prediction horizons of $k = 1$ and $k = 4$ quarters, respectively,
  for the first quarter of 1985 to the second quarter of 2011.  The
  means are computed on instances when the observation is smaller than
  0.10 (GDP) or larger than 0.98 (inflation) only.  For each variable
  and column, the lowest value is shown in bold.  \label{tab:cs2}}

\centering
 
\begin{tabular}{l@{\hskip 1.25cm}cc@{\hskip 1.25cm}cc}  
\toprule
& \multicolumn{2}{c}{\hspace{-1cm}rCRPS} & \multicolumn{2}{c}{rLogS} \\
\midrule
& $k = 1$ & $k = 4$ & $k = 1$ & $k = 4$ \\
\midrule
GDP Growth \\
\midrule
AR	   & \bf{0.654} & \bf{0.870} & \bf{1.626} & \bf{2.010} \\
AR-TVP-SV  & 0.659	    & 0.970	     & 2.016	      & 3.323 \\
VAR	   & 0.827	    & 0.924	     & 2.072	      & 2.270 \\
VAR-TVP-SV & 0.798	    & 0.978	     & 2.031	      & 2.409 \\
\midrule
Inflation \\
\midrule
AR	   & 0.214	    & 0.157	     & 0.484	      & \bf{0.296} \\
AR-TVP-SV  & 0.236	    & 0.179	     & 0.619	      & 0.327 \\
VAR	   & \bf{0.203} & \bf{0.147} & \bf{0.424} & 0.317 \\
VAR-TVP-SV & 0.302	    & 0.247	     & 0.950	      & 0.849 \\
\bottomrule
\end{tabular}
 
\end{table}

\subsection{Results}  \label{sec:results}

To compare the predictive performance of the four forecasting models,
Table~\ref{tab:cs1} shows the mean CRPS and LogS over the evaluation
period.  For the LogS, we follow extant practice in the economic
literature and employ the quadratic approximation proposed by
\citet{AdolfsonEtAl2007}.  Specifically, we find the mean,
$\hat{\mu}_F$, and variance, $\hat{\sigma}^2_F$, of a sample
$\hat{x}_1, \dots, \hat{x}_m$, where $\hat{x}_i$ is a random number
drawn from the $i$th mixture component of the posterior predictive
distribution \eqref{eq:pred.distr}, and compute the logarithmic score
under the assumption of a normal predictive distribution with mean
$\hat{\mu}_F$ and variance $\hat{\sigma}^2_F$.\footnote{We believe
that there are more efficient and more theoretically principled ways
of approximating the LogS in Bayesian settings.  However, these
considerations are beyond the scope of the paper, and we leave them to
future work.  Here, we use the quadratic approximation based on a
sample.  This very nearly corresponds to replacing the LogS by the
proper Dawid-Sebastiani score \citep[DSS;][]{DawidSebastiani1999,
GneitingRaftery2007}, which for a predictive distribution $F$ with
mean $\mu_F$ and finite variance $\sigma_F^2$ is given by
\[
\textnormal{DSS}(F,y) = 2 \log\sigma_F + \frac{(y-\mu_F)^2}{\sigma_F^2}. 
\]
The quadratic approximation is infeasible for the CL and CSL scoring
rules, as it then leads to improper scoring rules; see Appendix
\ref{sec:AppendixA}.}  To compute the CRPS and the threshold-weighted
CRPS, we use the numerical methods proposed by
\citet{GneitingRanjan2011}.

The relative predictive performance of the forecasting models is
consistent across the two variables and the two proper scoring rules.
The AR-TVP-SV model has the best predictive performance and
outperforms the baseline AR model. The $p$-values for the respective two-sided Diebold-Mariano tests range from 0.00 to 0.06, except for the LogS for GDP growth at a prediction horizon of $k = 4$ quarters, where the $p$-value is 0.37.
However, the VAR models fail to
outperform the simpler AR models.  As we do not impose sparsity
constraints on the parameters of the VAR models, this is likely due to
overly complex forecasting models and overfitting, in line with
results of \citet{HolzmannEulert2014} and \citet{ClarkRavazzolo2014}
in related economic and financial case studies.

To relate to the forecaster's dilemma, we restrict attention to
extremes events.  For GDP growth, we consider quarters with observed
growth less than $r = 0.1$ only.  For inflation, we restrict attention
to high values in excess of $r = 0.98$.  In either case, this
corresponds to using about 10\% of the observations.  Table
\ref{tab:cs2} shows the results of restricting the computation of the
mean CRPS and the mean LogS to these observations only.  For both GDP
growth and inflation, the baseline AR model is considered best, and
the AR-TVP-SV model appears to perform poorly.  These restricted
scores thus result in substantially different rankings than the proper
scoring rules in Table~\ref{tab:cs1}, thereby illustrating the
forecaster's dilemma.  Strikingly, under the restricted assessment all
four models seem less skillful at predicting inflation in the current
quarter than four quarters ahead.  This is a counterintuitive result
that illustrates the dangers of conditioning on outcomes and should be
viewed as a further manifestation of the forecaster's dilemma.

\begin{table}[t]

\caption{Mean threshold-weighted CRPS for probabilistic forecasts of
  GDP growth and inflation in the U.S.~at prediction horizons of $k =
  1$ and $k = 4$ quarters, respectively, under distinct weight
  functions, for the first quarter of 1985 to the second quarter of
  2011.  For each variable and column, the lowest value is shown in
  bold.  \label{tab:cs3}}
  
\centering
 
\begin{tabular}{l@{\hskip 1.25cm}cc@{\hskip 1.25cm}cc}  
\toprule
& \multicolumn{4}{c}{twCRPS} \\
\midrule
& $k = 1$ & $k = 4$ & $k = 1$ & $k = 4$ \\
\midrule
GDP Growth 
& \multicolumn{2}{c}{\hspace{-1.25cm}$w_{\textnormal{I}}(z) = \one \{ z \leq 0.1 \}$} 
& \multicolumn{2}{c}{\hspace{-0.50cm}$w_{\textnormal{G}} = 1 - \Phi( z \, | \, 0.1, 1)$} \\
\midrule
AR	   & 0.062	    & 0.068	     & 0.053	      & 0.057 \\
AR-TVP-SV  & \bf{0.052} & \bf{0.062} & \bf{0.048} & 0.055 \\
VAR	   & 0.062	    & \bf{0.062} & 0.054	      & \bf{0.054} \\
VAR-TVP-SV & 0.059	    & 0.080	     & 0.053	      & 0.065 \\
\midrule
Inflation 
& \multicolumn{2}{c}{\hspace{-1.25cm}$w_{\textnormal{I}}(z) = \one \{ z \geq 0.98 \}$} 
& \multicolumn{2}{c}{\hspace{-0.25cm}$w_{\textnormal{G}} = \Phi( z \, | \, 0.98, 1)$} \\
\midrule
AR	   & 0.026	    & 0.032	     & 0.068	      & 0.075 \\
AR-TVP-SV  & \bf{0.018} & \bf{0.018} & \bf{0.059} & \bf{0.065} \\
VAR	   & 0.027	    & 0.033	     & 0.072	      & 0.081 \\
VAR-TVP-SV & 0.022	    & 0.037	     & 0.067	      & 0.081 \\
\bottomrule
\end{tabular}

\end{table} 

In Table \ref{tab:cs3} we show results for the proper twCRPS under
weight functions that emphasize the respective region of interest.
For both variables, this yields rankings that are similar to those in
Table \ref{tab:cs1}. However, the $p$-values for binary comparisons with two-sided Diebold-Mariano tests generally are larger than those under the unweighted CRPS.
The AR-TVP-SV model is predominantly the best,
and the current quarter forecasts are deemed more skillful than those
four quarters ahead.

\section{Discussion}  \label{sec:discussion}

We have studied the dilemma that occurs when forecast evaluation is
restricted to cases with extreme observations, a procedure that
appears to be common practice in public discussions of forecast
quality.  As we have seen, under this practice even the most skillful
forecasts available are bound to be discredited when the
signal-to-noise ratio in the data generating process is low.  Key
examples might include macroeconomic and seismological predictions.
In such settings it is important for forecasters, decision makers,
journalists, and the general public to be aware of the forecaster's
dilemma.  Otherwise, charlatanes might be given undue attention and
recognition, and critical societal decisions could be based on
misguided predictions.

We have offered two complementary explanations of the forecaster's
dilemma.  From the joint distribution perspective of Section
\ref{sec:framework} stratifying by, and conditioning on, the realized
value of the outcome is problematic in forecast evaluation, as
theoretical guidance for the interpretation and assessment of the
resulting conditional distributions is unavailable.  In contrast
stratifying by, and conditioning on, the forecast is unproblematic.
From the perspective of proper scoring rules in
\ref{sec:understanding}, restricting the outcome space corresponds to
the multiplication of the scoring rule by an indicator weight
function, which renders any proper score improper, with an explicit
hedging strategy being available.

Arguably the only remedy is to consider all available cases when
evaluating predictive performance.  Proper weighted scoring rules
emphasize specific regions of interest and facilitate interpretation.
Interestingly, however, the Neyman-Pearson lemma and our simulation
studies suggest that in general the benefits of using proper weighted
scoring rules in terms of power are rather limited, as compared to
using standard, unweighted scoring rules.  Any potential advantages
vanish under weight functions with increasingly extreme threshold
values, where the finite sample behavior of Diebold-Mariano tests
depends on the tail properties of the forecast distributions only.

When evaluating probabilistic forecasts with emphasis on extremes, one
could also consider functionals of the predictive distributions, such
as the induced probability forecasts for binary tail events, as
utilized in a recent comparative study by \citet{WilliamsEtAl2014}.
Another option is to consider the induced quantile forecasts, or
related point summaries of the (tails of the) predictive
distributions, at low or high levels, say $\alpha = 0.975$ or $\alpha
= 0.99$, as is common practice in financial risk management, both for
regulatory purposes and internally at financial institutions
\citep{McNeilEtAL2015}.  In this context, \citet{HolzmannEulert2014}
studied the power of Diebold-Mariano tests for quantile forecasts at
extreme levels, and \citet{FisslerEtAl2015} raise the option of
comparative backtests of Diebold-Mariano type in banking regulation.
\citet{EhmEtAl2015} propose decision theoretically principled, novel
ways of evaluating quantile and expectile forecasts.

Variants of the forecaster's dilemma have been discussed in various
strands of literature.  Centuries ago, \citet{Bernoulli1713} argued
that even the most foolish prediction might attract praise when a rare
event happens to materialize, referring to lyrics by \citet{Owen1607}
that are quoted in the front matter of our paper.

\citet{Tetlock2005} investigated the quality of probability forecasts
made by human experts for U.S.~and world events.  He observed that
while forecast quality is largely independent of an expert's political
views, it is strongly influenced by how a forecaster thinks.
Forecasters who ``know one big thing'' tend to state overly extreme
predictions and, therefore, tend to be outperformed by forecasters who
``know many little things''.  Furthermore, \citet{Tetlock2005} found
an inverse relationship between the media attention received by the
experts and the accuracy of their predictions, and offered
psychological explanations for the attractiveness of extreme
predictions for both forecasters and forecast consumers.  Media
attention might thus not only be centered around extreme events, but
also around less skillful forecasters with a tendency towards
misguided predictions.

\citet{DenrellFang2010} reported similar observations in the context
of managers and entrepreneurs predicting the success of a new product.
They also studied data from the Wall Street Journal Survey of Economic
Forecasts, found a negative correlation between the predictive
performance on a subset of cases with extreme observations and
measures of general predictive performance based on all cases, and
argued that accurately predicting a rare and extreme event actually is
a sign of poor judgment.  Their discussion was limited to point
forecasts, and the suggested solution was to take into account all
available observations, much in line with the findings and
recommendations in our paper.

\bibliographystyle{ims}
\bibliography{bibliography}

\section*{Acknowledgments}

The support of the Volkswagen Foundation through the project
`Mesoscale Weather Extremes --- Theory, Spatial Modeling and Prediction
(WEX-MOP)' is gratefully acknowledged.  Sebastian Lerch also
acknowledges support by the Deutsche Forschungs\-gemeinschaft through
Research Training Group 1953, and Tilmann Gneiting and Sebastian Lerch
are grateful for support by the Klaus Tschira Foundation.  The initial
impetus for this work stems from a meeting with Jeff Baars, Cliff Mass
and Adrian Raftery at the University of Washington, where Jeff Baars
presented a striking meteorological example of what we here call the
forecaster's dilemma.  We are grateful to our colleagues for the
inspiration.  We thank Norbert Henze for insightful comments on
initial versions of our simulation studies, and Alexander Jordan for
suggesting the simulation setting in Section \ref{sec:sim2b}.  We also
are grateful to Richard Chandler for pointing us to the Neyman-Pearson
connection and the paper by \citet{FeuervergerRahman1992}.

\newpage

\appendix

\section{Impropriety of quadratic approximations of weighted
  logarithmic scores}  \label{sec:AppendixA}
 
Let $F$ be a predictive distribution with mean $\mu_F$ and standard
deviation $\sigma_F$.  As regards the conditional likelihood (CL)
score \eqref{eq:CL}, the quadratic approximation is given by
\[
\textnormal{CL}^q(F,y) = 
- w(y) \log \! \left( \frac{\phi(y|F)}{\int w(x) \phi(x|F) \, \dd x} \right) \! ,
\]
where $\phi(\cdot|F)$ denotes a normal density with mean $\mu_F$ and
standard deviation $\sigma_F$, respectively.  Let
\[
c_F = \int w(x)\phi(x|F) \, \dd x, \qquad 
c_G = \int w(x)\phi(x|G) \, \dd x, \qquad
c_g = \int w(x)g(x) \, \dd x,
\]
and recall that the Kullback-Leibler divergence between two probability densities $u$ and $v$
is given by 
\[
K(u,v) = \int u(x) \log \! \left( \frac{u(x)}{v(x)} \right) \dd x. 
\]
Assuming that CL$^q$ is proper, it is true that  
\begin{align*}
\myE_G & (\textnormal{CL}^q(F,Y) - \textnormal{CL}^q(G,Y)) \\ 
& = c_g \left[ K \! \left( \frac{w(y)g(y)}{c_g}, \frac{w(y)\phi(y|F)}{c_F} \right) 
             - K \! \left( \frac{w(y)g(y)}{c_g}, \frac{w(y)\phi(y|G)}{c_G} \right) \right]
\end{align*}
is non-negative.  Let $G$ be uniform on $[-\sqrt 3,\sqrt 3]$ so that
$\mu_G = 0$ and $\sigma_G = 1$, and let $w(y) = \one \{ y \geq 1 \}$.
Denoting the cumulative distribution function of the standard normal
distribution by $\Phi$, we find that
\begin{align*}
K & \! \left( \frac{w(y)g(y)}{c_g}, \frac{w(y)\phi(y|F)}{c_F} \right) 
    - K \! \left( \frac{w(y)g(y)}{c_g}, \frac{w(y)\phi(y|G)}{c_G}\right) \\
& = \log \! \left( \sigma_F \frac{1 - \Phi((1-\mu_F)/\sigma_F)}{1 - \Phi(1)} \right) 
    + \frac{3(\sqrt{3}-1)\mu_F^2 - 6 \mu_F + (3\sqrt{3} - 1)(1 - \sigma_F^2)}{6(\sqrt{3}-1)\sigma_F^2}, 
\end{align*}
which is strictly negative in a neighborhood of $\mu_F = 1.314$ and
$\sigma_F = 0.252$, for the desired contradiction.  Therefore, CL$^q$
is not a proper scoring rule.
 
As regards the censored likelihood (CSL) score \eqref{eq:CSL}, the
quadratic approximation is
\[
\textnormal{CSL}^q(F,y) = 
- w(y) \log \! \left( \phi(y|F) \right) 
- (1-w(y)) \log \! \left( 1- \int w(z)\phi(z|F) \, \dd z \right) \! . 
\]
Under the same choice of $w$, $F$, and $G$ as before, we find that
\begin{align*}
\myE_G & (\textnormal{CSL}^q(F,Y) - \textnormal{CSL}^q(G,Y)) \\ 
& = \frac{\sqrt{3} - 1 }{2\sqrt 3} \log \sigma_F - \frac{\sqrt{3} + 1}{2\sqrt{3}} 
    \log \! \left( \frac{\Phi((1-\mu_F)/\sigma_F)}{\Phi(1)} \right) \\
& \hspace{15mm} + \frac{3(\sqrt{3}-1)\mu_F^2 - 6\mu_F + (3\sqrt{3} - 1)(1 - \sigma_F^2)}{12 \sqrt{3} \, \sigma_F^2},
\end{align*}
which is strictly negative in a neighborhood of $\mu_F = 0.540$ and
$\sigma_F = 0.589$.  Therefore, CSL$^q$ is not a proper scoring rule.

\section{Online supplement: Media attention on extreme events}  \label{sec:AppendixB}

\begin{table}[p]

\begin{threeparttable}

\footnotesize
  
\centering
 
\caption{Media coverage illustrating the focus on extreme events in
  public discussions of the quality of forecasts.  The sources were
  accessed August 8, 2015.}
 
\begin{tabular}{lll} 
\toprule
Year & Headline & Source  \\
\midrule
2008 & Dr.~Doom & The New York Times\footnotemark[1] \\
2009 & How did economists get it so wrong? & The New York Times\footnotemark[2] \\
2009 & He told us so & The Guardian\footnotemark[3] \\
2010 & Experts who predicted US economy crisis see recovery & Bloomberg\footnotemark[4] \\
2010 & An exclusive interview with Med Yones - The expert who & CEO Q Magazine\footnotemark[5] \\
     & predicted the financial crisis & \\
2011 & A seer on banks raises a furor on bonds & The New York Times\footnotemark[6] \\
2013 & Meredith Whitney redraws 'map of prosperity' & USA Today\footnotemark[7] \\
\midrule
2007 & Lessons learned from Great Storm & BBC\footnotemark[8] \\
2011 & Bad data failed to predict Nashville Flood & NBC\footnotemark[9] \\
2012 & Bureau of Meteorology chief says super storm `just blew up  & The Courier-Mail\footnotemark[10] \\
     & on the city'  & \\
2013 & Weather Service faulted for Sandy storm surge warnings & NBC\footnotemark[11] \\
2013 & Weather Service updates criteria for hurricane warnings, & Washington Post\footnotemark[12] \\
     & after Sandy criticism  & \\
2015 & National Weather Service head takes blame for forecast & NBC\footnotemark[13] \\
     & failures & \\
\midrule
2011 & Italian scientists on trial over L'Aquila earthquake & CNN\footnotemark[14] \\
2011 & Scientists worry over `bizarre' trial on earthquake & Scientific American\footnotemark[15] \\
     & prediction & \\
2012 & L'Aquila ruling: Should scientists stop giving advice? & BBC\footnotemark[16] \\
\bottomrule
\end{tabular}
 
\vspace{0.2cm}

\begin{tablenotes}
\item[1]{\url{http://www.nytimes.com/2008/08/17/magazine/17pessimist-t.html?pagewanted=all}}
\item[2]{\url{http://www.nytimes.com/2009/09/06/magazine/06Economic-t.html?_r=1&pagewanted=all}}
\item[3]{\url{http://www.guardian.co.uk/business/2009/jan/24/nouriel-roubini-credit-crunch}}
\item[4]{\url{http://www.bloomberg.com/apps/news?pid=conewsstory&refer=conews&tkr=K:US&sid=asziFnEsJSos}}
\item[5]{\url{http://www.ceoqmagazine.com/whopredictedfinancialcrisis/index.htm}}
\item[6]{\url{http://www.nytimes.com/2011/02/08/business/economy/08whitney.html?pagewanted=all&_r=0}}
\item[7]{\url{http://www.usatoday.com/story/money/business/2013/06/05/meredith-whitney-book-interview/2384905/}}
\item[8]{\url{http://news.bbc.co.uk/2/hi/science/nature/7044050.stm}}
\item[9]{\url{http://www.nbc15.com/weather/headlines/January_13_Report_Bad_Data_Failed_To_Predict_Nashville_Flood_113450314.html}}
\item[10]{\url{http://www.couriermail.com.au/news/queensland/bureau-of-meteorology-under-fire-after-a-weekend-of-wild-weather-and-storms-in-queensland-left-many-unprepared/story-e6freoof-1226519213928}}
\item[11]{\url{http://www.nbcnewyork.com/news/local/Sandy-Report-Weather-Storm-Surge-Warnings-207545031.html}}
\item[12]{\url{http://www.washingtonpost.com/blogs/capital-weather-gang/wp/2013/04/04/weather-service-changes-criteria-for-hurricane-warnings-after-sandy-criticism/}}
\item[13]{\url{ http://www.nbcnews.com/storyline/blizzard-15/national-weather-service-head-takes-blame-forecast-failures-n294701}}
\item[14]{\url{http://articles.cnn.com/2011-09-20/world/world_europe_italy-quake-trial_1_geophysics-and-vulcanology-l-aquila-seismic-activity?_s=PM:EUROPE}}
\item[15]{\url{http://www.scientificamerican.com/article.cfm?id=trial-such-as-that-star}}
\item[16]{\url{http://www.bbc.co.uk/news/magazine-20097554}}
\end{tablenotes}

\end{threeparttable}

\end{table}

\end{document}